\documentclass[11pt,a4paper]{article}
\pdfoutput=1
\usepackage{jheppub}

\makeatletter
\def\@fpheader{\relax}
\makeatother

\usepackage{graphicx}

\newcommand{\be}{\begin{equation}}
\newcommand{\beq}{\begin{equation}}
\newcommand{\en}{\end{equation}}
\newcommand{\eeq}{\end{equation}}
\newcommand{\bea}{\begin{eqnarray}}
\newcommand{\ena}{\end{eqnarray}}

\newcommand{\Z}{\mathbb{Z}}

\newcommand{\VEV}[1]{\left\langle #1 \right\rangle}

\newcommand{\Tr}{{\rm Tr\,}}
\newcommand{\badj}{\beta_{\mathrm{adj}}}
\newcommand{\bfund}{\beta_{\mathrm{fund}}}
\newcommand{\chif}{\chi_{\rm Pf}}
\newcommand{\chia}{\chi_{\rm Pa}}
\newcommand{\Msc}{am_{0^{++}}}
\newcommand{\Mtn}{am_{2^{++}}}
\newcommand{\Mst}{a\sqrt{\sigma}}

\affiliation[a]{
College of Science, Swansea University, Singleton Park, Swansea SA2 8PP, UK
}

\affiliation[b]{
PH-TH, CERN, CH-1211 Geneva 23, Switzerland
}
\affiliation[c]{
School of Computing and Mathematics \& Centre for Mathematical Science, Plymouth University, Plymouth PL4 8AA, UK
}

\affiliation[d]{
  Higgs Centre for Theoretical Physics, SUPA, School of Physics and Astronomy, University of Edinburgh, Edinburgh EH9 3JZ, UK
}

\preprint{CERN-PH-TH/2013-215, Edinburgh 2013/12}

\title{Infrared conformality and bulk critical points: SU(2) with heavy adjoint quarks}

\author[a]{Biagio Lucini}
\author[b,c]{Agostino Patella}
\author[c]{Antonio Rago}
\author[d]{Enrico Rinaldi}

\emailAdd{b.lucini@swansea.ac.uk, agostino.patella@plymouth.ac.uk, antonio.rago@plymouth.ac.uk, e.rinaldi@sms.ed.ac.uk}

\abstract{
The lattice phase structure of a gauge theory can be a serious
obstruction to Monte Carlo studies of its continuum behaviour.
This issue is particularly delicate when numerical studies
are performed to determine whether a theory is in a (near-)conformal phase.
In this work we investigate the heavy mass limit of the SU($2$) gauge
theory with $N_f=2$ adjoint fermions and its lattice phase diagram,
showing the presence of a critical point ending a line of first order
bulk phase transition. The
relevant gauge observables and the low--lying spectrum are monitored
in the vicinity of the critical point with very good control over
different systematic effects. The scaling properties of masses and
susceptibilities open the possibility that the effective theory at
criticality is a scalar theory in the universality class of the 
four--dimensional Gaussian model.
This behaviour is clearly different from what is
observed for SU(2) gauge theory with two dynamical adjoint fermions, whose
(near-)conformal numerical signature is hence free from strong--coupling
bulk effects.
}

\begin{document}

\maketitle

\section{Introduction}
\label{sec:introduction}
Despite the recent identification of a light Higgs boson at the
LHC~\cite{atlas,cms}, unveiling the mechanism of electroweak symmetry breaking
is still an open problem in theoretical particle physics. Among the
possibilities still on the table, the suggestion that a novel strong
interaction displaying
confinement~\cite{Weinberg1975gm,Susskind1978ms} and an anomalous
dimension of the chiral condensate of order one~\cite{Holdom1984sk,Yamawaki1985zg,Appelquist1986an} 
can be tested quantitatively with lattice simulations. In the quest
for a theory that could realise concretely this scenario, several
gauge theory models, with matter in the fundamental or in a two--index
representation and various flavour and colour content have been
studied with Monte Carlo methods
(see~\cite{DelDebbio2010zz,Neil2012cb,Giedt:2012it} for recent
reviews), with the space of parameters narrowed down using
analytical input~\cite{Dietrich2006ck,Dietrich2006cm}. 

A large anomalous dimension is expected to arise near the onset of the
conformal window. Hence, to unambiguously ascertain conformality for a
gauge theory, one needs to be able to robustly determine whether the
theory has or is near to an infrared (IR) fixed point. Recent
Monte Carlo studies have shown that the identification of IR fixed
points on the lattice is not straightforward, since in numerical
simulations the system has a finite size (while conformality would require
infinite distances to be explored) and fermions have a finite mass (the conformal
limit being for massless matter fields). In addition, lattice
simulations are performed at fixed cutoff and for particular choices
of the discretised action. Because of these unavoidable complications,
there is wide consensus that, in lattice simulations aimed at
ascertaining (near-)conformality of a gauge theory, evidences based on different approaches
and different techniques need to be collected before one can exclude
spurious lattice signatures being mistaken for genuine IR fixed points
in the continuum. In the last few years, thanks to the joint effort of
various groups, this programme has been carried out for
SU(2) gauge theory with two adjoint Dirac fermions, for which the scaling of the
spectrum~\cite{DelDebbio2009fd,Catterall2009sb,Lucini2009an,DelDebbio2010hx,DelDebbio2010hu} and
the behaviour of the coupling constant under RG
flow~\cite{Hietanen2009az,Bursa2009we,DeGrand2011qd,Catterall2011zf}
strongly suggest IR conformality (see
also~\cite{Catterall2007yx,DelDebbio2008zf,Hietanen2009zz} for earlier 
numerical investigations). More recent lattice studies of the theory
are focused on controlling systematic
effects~\cite{Bursa2011ru,Karavirta2011mv,DelDebbio2011kp,Bennett:2012ch}
and on precise measurements of the anomalous
dimension~\cite{Giedt2012rj,Patella2012da}. However the persistence of
the IR fixed point in the continuum limit is still under
investigation.

Lattice simulations in the (near-)conformal regime are made more
difficult by the lack of experimental guidance. For this reason, all
possible sources of uncertainties and ambiguities need to be carefully
analysed. In this paper, we study the possibility that numerical
indications of conformality in SU(2) gauge theory with two Dirac flavours in the 
adjoint representation are in fact due to the presence of a second
order transition point in a system related to this theory. Namely we
refer to a SU(2) gauge theory with mixed fundamental--adjoint
action~\cite{Bhanot:1981eb}, which the aforementioned theory with
dynamical fermions reduces to at leading non--trivial order in the
hopping parameter expansion. This is a
potential effect that has not been considered before in the
literature. For its investigation, we will compare the scaling of the
spectrum in the gluonic sector found for SU(2) with two adjoint Dirac
fermions with the scaling of the pure gauge spectrum of the mixed
action system near its quantum critical point. In doing so, we shed
some light on the nature of this point, solving some controversies
in the earlier literature~\cite{Gavai:1994bt,Mathur:1994nt,Stephenson:1996ny,Gavai:1996yc,Datta:1996pi,Grady:2004fi}.  

The rest of the paper is organised as follows. After
introducing the system we have investigated and elucidating its
relationship with SU(2) with two adjoint Dirac fermions
(Sect.~\ref{sec:model}), in Sect.~\ref{sec:phase-diagram}
we present our results on the location of the critical point in the bare
coupling plane. These are obtained from the study of plaquette differences in the two
coexisting vacua along the first order phase transition line ending in
it. We also show results for the susceptibility of the
plaquette. Sect.~\ref{sec:spectr-meas} reports on our data for the
spectrum, whose scaling properties are investigated in
Sect.~\ref{sec:scal-prop}. Our findings are then summarised in
Sect.~\ref{sect:conclusions}.
Preliminary results of our investigation have been reported
in~\cite{Rinaldi:2012vb}.  

\section{The model}
\label{sec:model}
In the Wilson discretisation of fermions, the lattice Dirac
operator for a single fermion species of mass $a m$ (in lattice
units) transforming in the representation $\mathrm{R}$ of the gauge
group is given by 
\begin{eqnarray}
\nonumber
 M_{\alpha \beta}(ij) &=& (a m+4r)  \delta_{ij}
  \delta_{\alpha \beta} - \frac{1}{2} \left[\left(r - \gamma_{\mu}\right)_{\alpha
      \beta}U^{\mathrm{R}}_{\mu}(i) \delta_{i,j-\hat{\mu}} \right.\\
    &&\quad+ \left. \left(r  +
      \gamma_{\mu}\right)_{\alpha \beta} \left(U^{\mathrm{R}}_{\mu}(i)\right)^{\dag}\delta_{i,j+\hat{\mu}}
  \right] \ ,
\end{eqnarray}
where $i$ and $j$ are lattice site indices, $\alpha$ and $\beta$ are
Dirac indices, $\mu$ is an Euclidean direction and the $\gamma$
matrices are formulated in Euclidean space. 
$U^{\mathrm{R}}_{\mu}(i)$ is the link variable in the representation
$\mathrm{R}$ of the gauge group SU($N_c$).  The path integral of a theory with
$N_f$ flavours transforming in the representation $\mathrm{R}$ is then
given by
\begin{eqnarray}
 Z = \int \left( {\cal D} U_{\mu}(i)\right) (\det M(U_{\mu}))^{N_{\mathrm{f}}}
  e^{-S_{\mathrm{F}}} \ ,
\end{eqnarray}
where $S_{\mathrm{F}}$ is the gauge action, which for simplicity will
be taken as the Wilson plaquette action:
\begin{equation}
  \label{eq:action-lattice-fund}
    S_{\mathrm{F}} =  \bfund \sum_{i,\mu > \nu} \left( 1 - \frac{1}{N_c} \mbox{Re} \ \Tr_F
      \left( U_{\mu \nu}(i) \right) \right) \ .
\end{equation}
Here, $U_{\mu \nu}(i)$ the plaquette in the $(\mu,\nu)$--plane from
point $i$ and $\bfund = 2 N_c/g^2$, with $g$ the bare coupling. The
sum over all the points $i$ is done over the four--dimensional lattice
$L^3 \times T$. $\Tr_F$ is the trace operator defined in the
fundamental representation of the SU($N_c$) gauge group. For reasons
that will be clear below, we call this the fundamental action (where
fundamental refers to the fact that the plaquette is in the
fundamental representation).

For large bare quark mass, $\det M$ can be expanded in powers of the hopping
parameter $\kappa = [2(a m + 4r)]^{-1}$. At the leading non--trivial
order, this gives
\begin{eqnarray}
 Z = \int \left( {\cal D} U_{\mu}(i)\right) e^{-S_{\mathrm{eff}}} \ ,
\end{eqnarray}
with 
\begin{equation}
\label{eq:variantgen}
S_{\mathrm{eff}} = S_{\mathrm{F}} + S_{\mathrm{R}} 
\end{equation}
and (up to irrelevant constants)
\begin{eqnarray}
\label{eq:sr}
S_R = \tilde{\beta}_{\mathrm{R}} \sum_{i,\mu > \nu}
    \left( 1 - \frac{1}{d_{\mathrm{R}}} \mbox{Re} \ \Tr_{\mathrm R}
      \left( U_{\mu \nu}(i) \right) \right) \ ,
\end{eqnarray}
where $\Tr_{\mathrm R}$ is the trace in the representation $\mathrm{R}$,
$d_{\mathrm{R}}$ the dimension of that representation and
\begin{equation}
\label{eq:betatilde}
\tilde{\beta}_{\mathrm{R}} = 8 \kappa^4  d_{\mathrm R} \left (1 + 2 r^2 -
  r^4 \right) \ .
\end{equation}
Eqs.~(\ref{eq:variantgen}-\ref{eq:betatilde}) show that at high bare mass the dynamical system
is approximated by a gauge system with a mixed action, i.e. with an
action that, in addition to the fundamental Wilson term, has a coupling to the plaquette in the
representation $\mathrm{R}$ governed by the mass of the fermions
(assumed to be large). These variant actions are known to have a
non-trivial phase structure in the plane of the couplings (see e.g.~\cite{Caneschi:1981ik}). 

If we specialise our derivation to the adjoint representation,
Eq.~(\ref{eq:variantgen}) becomes a particular case of the mixed
fundamental--adjoint gauge action
\begin{equation}
  \label{eq:action-lattice}
  \begin{split}
    S & =  \bfund \sum_{i,\mu > \nu} \left( 1 - \frac{1}{N_c} \mbox{Re} \ \Tr_F
      \left( U_{\mu \nu}(i) \right) \right) \\ & + \badj \sum_{i,\mu > \nu}
    \left( 1 - \frac{1}{N_c^2 - 1} \mbox{Re} \ \Tr_A
      \left( U_{\mu \nu}(i) \right) \right) \ ,
  \end{split}
\end{equation}
where $\Tr_A$ is the trace in the adjoint representation (whose dimension
is $d_{A} = N_c^2 - 1$), related to the
trace in the fundamental representation $\Tr_F$ by $\Tr_A (U) =
|\Tr_F(U)|^2 - 1$. If we take SU(2) as the gauge group, the
action~(\ref{eq:action-lattice}) can be seen as a generalisation of
the high bare mass regime of SU(2) gauge theory with two adjoint Dirac
flavours.

For $N_c = 2$, simulations with the mixed fundamental--adjoint action
were already carried out in the early days of lattice gauge
theories~\cite{Bhanot:1981eb} and more recently in
Ref.~\cite{Gavai:1994bt,Mathur:1994nt,Gavai:1996yc,Datta:1996pi}. A  
mixed--action study of the finite temperature transition for SU(3) was
also considered in Ref.~\cite{Heller:1995bz}. These studies suggest
the existence of a second order phase transition point in the bare
coupling plane, which for SU(2) is attained at $\badj \approx 1.25$
and $\bfund \approx 1.22$. However, due to growing autocorrelation
times, which makes it difficult to study the system in a neighbourhood of
this critical point, this evidence has not been considered
conclusive~\cite{Grady:2004fi}. 

Due to the relationship between SU(2) with mixed fundamental--adjoint
action and SU(2) with two Dirac flavours in the adjoint
representation and to hints for near--conformality in the latter, it is
important to analyse carefully the physics of the pure gauge system in
its bare parameter space. In particular we explore the region around
$(\bfund = 1.22, \badj = 1.25)$ (see Sect.~\ref{sec:phase-diagram} for
details), to understand the signatures of a possible end--point
on the physical spectrum and to compare the behaviour of masses near it
with the scaling of masses in SU(2) with two adjoint Dirac flavours.

\begin{figure}[!h]
  \centering
  \includegraphics[width=0.6\textwidth]{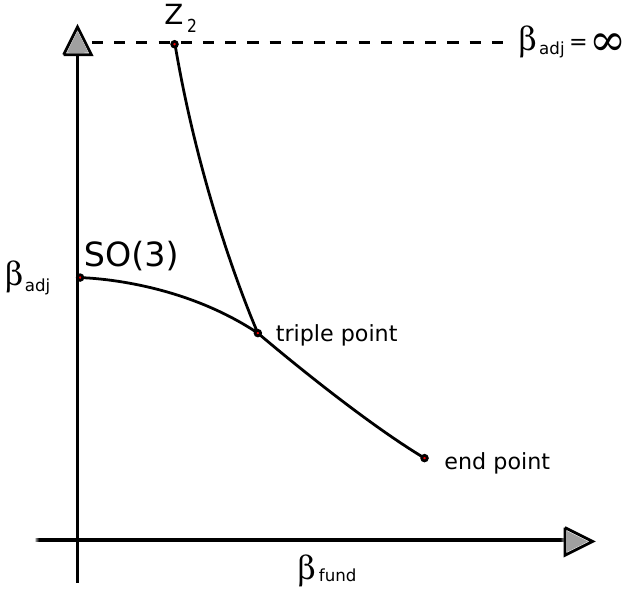}
  \caption{Sketch of the phase diagram for the lattice system defined
    by Eq.~(\ref{eq:action-lattice}). The lines of first order phase
    transitions are shown and are explained in Sect.~\ref{sec:phase-diagram}.}
  \label{fig:sketch-phase-diagram}
\end{figure}

\section{Phase diagram}
\label{sec:phase-diagram}
From previous works on small lattices~\cite{Bhanot:1981eb} and from
analytical arguments~\cite{Caneschi:1981ik}, it was known that the lattice
system described by Eq.~(\ref{eq:action-lattice}) had an interesting
non--trivial phase diagram. A sketch of the phase diagram is presented
in Fig.~\ref{fig:sketch-phase-diagram}.
The bare parameter space of the couplings ($\bfund$,$\badj$) contains
two first order phase transition lines that belong to theories in
different limits of the system: the $\Z_2$ gauge theory at $\badj =
\infty$ and the SO(3) gauge theory at $\bfund = 0$. These two lines
merge in a triple point at finite values of the couplings and continue as a single line
toward the fundamental Wilson axis $\badj=0$. This single line is
thought to be a line of bulk first order transition points that ends around
$\badj \approx 1.25$. Due to the possibility of this end--point being of
second order, it is necessary to carefully investigate the nature of
the transition on large volume lattices and at high statistics, which
is required by the large autocorrelation times of the system.
For our Monte Carlo simulations we
employ a biased Metropolis algorithm that has been proven to have an
heatbath--like efficiency~\cite{Bazavov:2005vr,Bazavov:2005zy}. This algorithm helped
us reduce autocorrelations with respect to the standard multi--hit
Metropolis on which earlier results are based\footnote{An alternative
algorithm with an efficiency similar to the one used in this work has
been proposed in~\cite{Vairinhos:2010ha}.}.

\begin{figure}[t]
  \centering
  \includegraphics[width=0.7\textwidth]{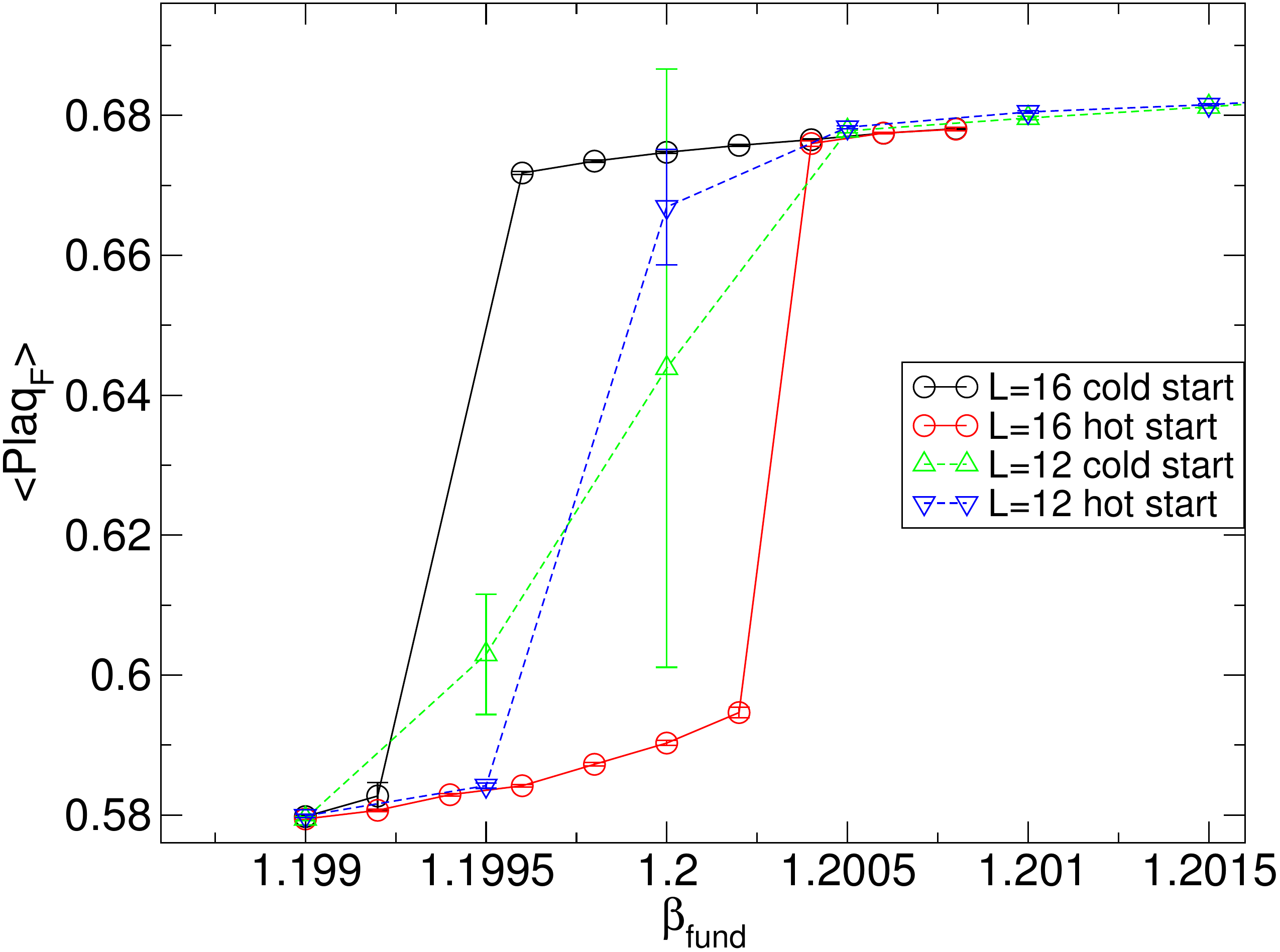}
  \caption{Fundamental plaquette expectation values for several $\bfund$
    couplings at $\badj = 1.275$ on two different volumes
    $L=12,16$. The hysteresis cycle, clearly visible on the larger
    volume, is hard to identify on the smaller one. A similar
    consistent picture holds for the expectation values of the modulus
    of Polyakov loops.} 
  \label{fig:plaqF-ba1275}
\end{figure}
We first checked for the presence of the bulk transition line on
hypercubic lattices as large as $L=T=40$ by simulating at $\badj >
1.25$. We monitored local observables such as the fundamental and
adjoint plaquettes and the Polyakov loops in the four directions,
together with their normalised susceptibilities. The
bulk transition manifests itself with a jump in the expectation value
of the plaquettes as the couplings are varied in the
(pseudo-)critical region. Moreover, in the same region a clear
hysteresis cycle in the 
plaquettes appears when Markov chains are started from random (hot) or
unit (cold) gauge configurations. The presence of metastable states
characterised by different values of the plaquette allows us to follow the bulk
transition line and estimate the location of its end--point, where the
plaquette gap disappears in the thermodynamic limit.

\begin{figure}[h]
  \centering
  \includegraphics[width=0.7\textwidth]{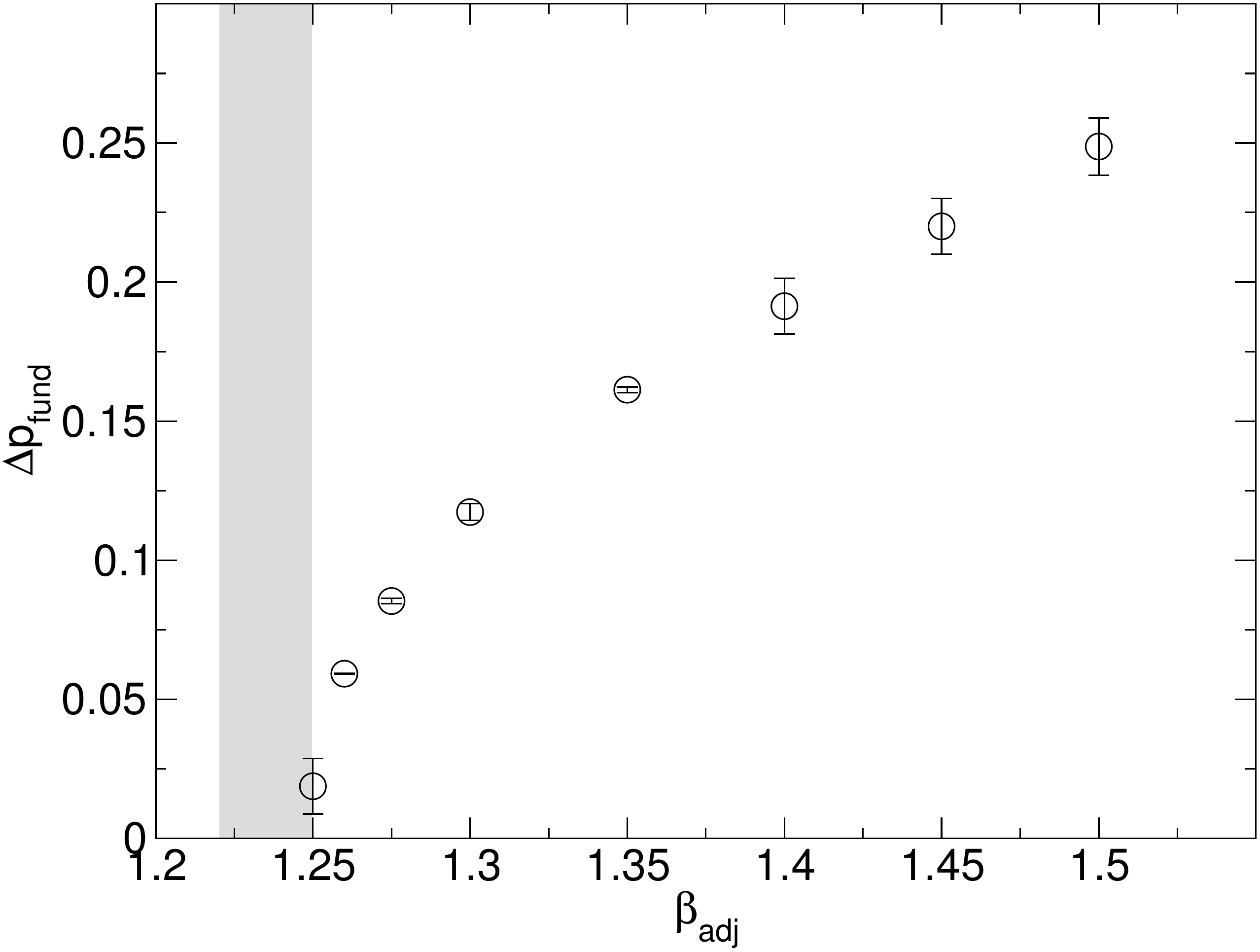}
  \caption{$\Delta p_{\rm fund}$ as defined in Eq.~(\ref{eq:latent-heat}). Also
    shown as a grey shaded area is the approximate position of the critical $\badj$ value
    at which $\Delta p_{\rm fund}$ is expected to vanish: $1.22 <
    \badj < 1.25$. A consistent
    result is found using the adjoint plaquette ($\Delta p_{\rm adj}$). The
    plotted points are reported in Tab.~\ref{tab:delta-p-f}.}
  \label{fig:latent-heat}
\end{figure}

The fundamental plaquette is shown in Fig.~\ref{fig:plaqF-ba1275} for
several $\bfund$ values in the pseudo--critical range at fixed
$\badj=1.275$. A similar behaviour is found for the vacuum expectation value 
of the modulus of the Polyakov loop. As expected, increasingly larger
lattice volumes are
requested to correctly identify the presence of the two metastable
states (and consequently of the hysteresis cycle) when $\badj$
approches $1.25$ from above. However, we noted that the transition to
the asymptotic regime is sharp, with values of relevant observables
approximately independent of $L$ and $T$ as soon as the small volume
regime is exited. On the largest volume used for this part of the
study, for which $L=T=40$, and with
more than 400000 measurements, we find that the plaquette gap between the
two vacua is non--zero at $\badj=1.25$, in contrast with the results at
smaller $L$ of Ref.~\cite{Gavai:1996yc}. A careful study of the plaquette
gap, defined as 
\begin{equation}
  \label{eq:latent-heat}
  \Delta p_{\rm fund} \; = \; \VEV{{\rm Plaq}_{F,1}} -
  \VEV{{\rm Plaq}_{F,2}}
  \ ,
\end{equation}
where the subscripts $1$ and $2$ refer to the distinct vacua at
couplings centered in the hysteresis loop, shows a definite
trend towards zero. The trend in $\Delta p_{\rm fund}$ is
plotted in Fig.~\ref{fig:latent-heat}, where, for each estimate, we
used the smallest volume where the first order nature of the
transition was manifest. Numerical values are summarised in
Tab.~\ref{tab:delta-p-f}. We note that the autocorrelation time
dramatically increases as one approaches the critical point (see
the $\tau_f$ and $\tau_a$ columns of Tab.~\ref{tab:susept-all}).  We were unable to give
an estimate of the autocorrelation time at $\badj = 1.25$, where a
large $L = 40$ lattice was required. Hence, on this latter
point the systematic error is not fully under control. For this reason
the point at $\badj = 1.25$ will not be considered in our subsequent
analyses. 
 
By interpolating both $\Delta p_{\rm fund}$
and the equivalent quantity $\Delta p_{\rm adj}$ for the adjoint
plaquette using polynomials, we identify a region $1.22 < \badj <
1.25$ where the plaquette gap vanishes. This is our first estimate for
the location of the end--point. This estimate will be
refined in the following using scaling analysis of the
susceptibilities and of the spectral observables.

\begin{figure}[ht]
  \centering
  \includegraphics[width=0.7\textwidth]{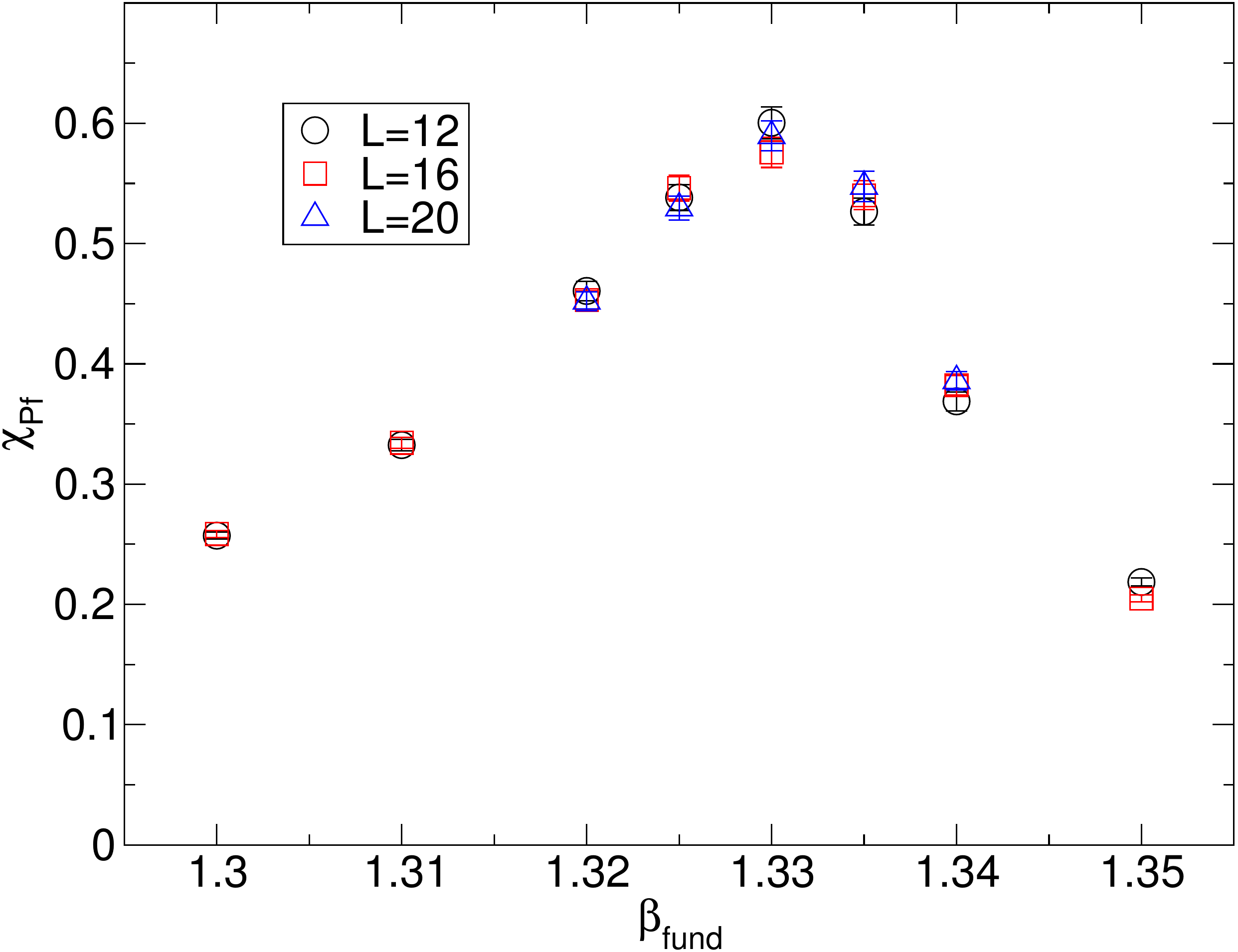}
  \caption{Lack of volume scaling in the fundamental plaquette susceptibility
    at $\badj = 1.10$. The peak location does not move up to volumes
    $L\times T=20^4$.}
  \label{fig:plaquette-susc}
\end{figure}

In the region below the approximate location of the end--point, we have
checked that the transition 
becomes a crossover, signalled by the lack of volume scaling 
in the fundamental and adjoint plaquette susceptibilities $\chif$,
$\chia$: the height and 
the location of peaks of the susceptibility (determined with a scan in
$\bfund$ at fixed $\badj$) are consistent across the
different volumes.  An example is reported in
Fig.~\ref{fig:plaquette-susc}. The location
of the peak can be followed in the  ($\bfund$,$\badj$) plane and
separates a strong coupling region at small $\bfund$ from 
a region closer to the weak coupling limit ($\bfund \rightarrow
\infty$). We summarise the maximum values for $\chif$ and $\chia$
along this crossover line in Tab.~\ref{tab:susept-all}. In the same
table we also report an estimate of the integrated autocorrelation
times for the fundamental and adjoint plaquettes.

\begin{figure}[ht]
  \centering
  \includegraphics[width=0.7\textwidth]{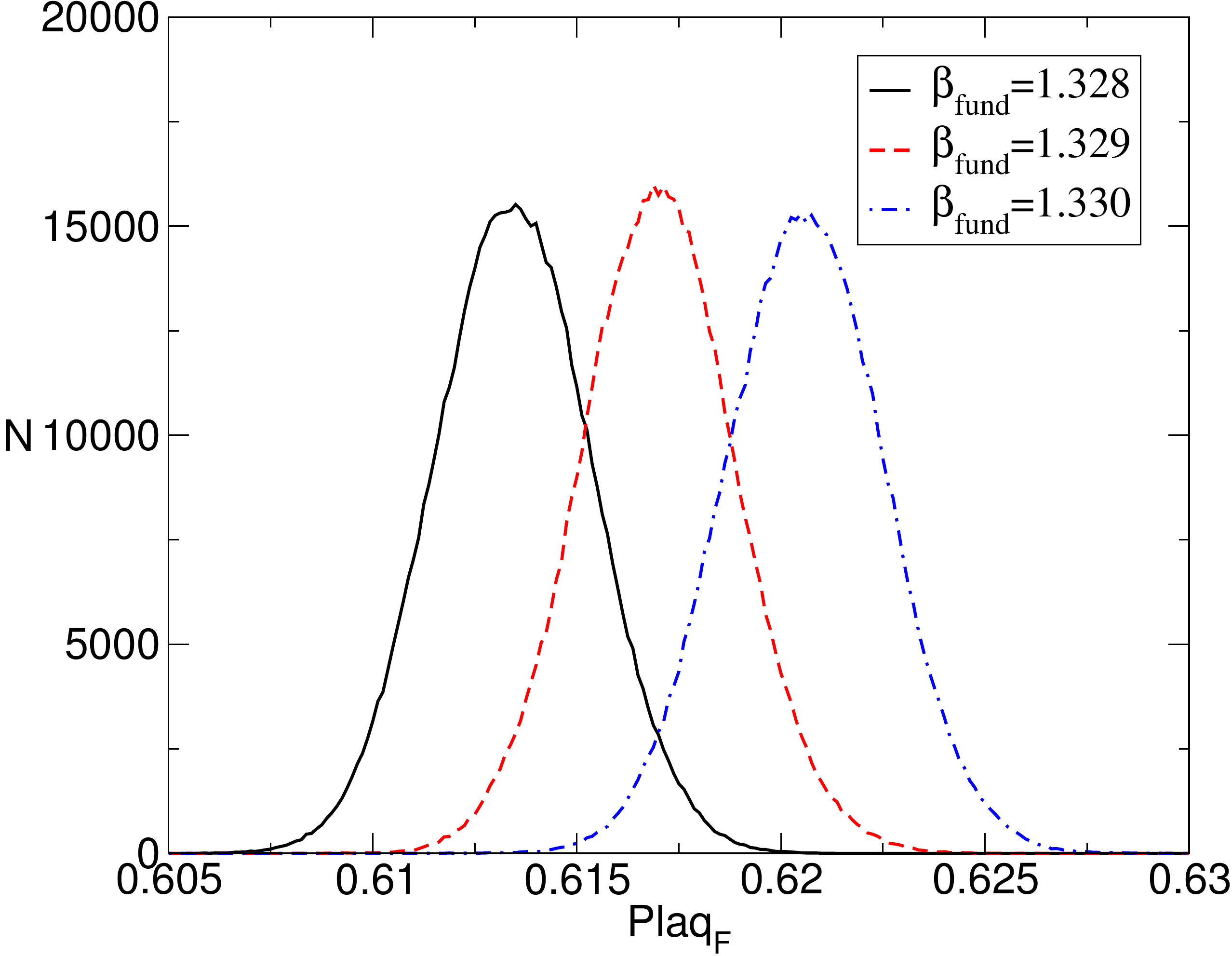}
  \caption{Fundamental plaquette distributions
    at $\badj = 1.10$ for $L\times T=20^4$. The overlap of different
    distributions is not enough to allow for a stable multi--histogram
  reweighting analysis of the susceptibility. Note that the distance
  between the points is much finer than the one in Fig.~\ref{fig:plaquette-susc}.}
  \label{fig:plaquette-distribution}
\end{figure}

The peak of the plaquette susceptibilities can be used to
give a new estimate of the end--point location. In a similar analysis,
the conventional way of proceeding is to use reweighting to locate the
maximum of susceptibilities. We have attempted this procedure, but
reweighting proves to be unviable due to the small overlap of the
plaquette distributions at neighbour $\bfund$ for lattices of the required
size. This situation is depicted in Fig.~\ref{fig:plaquette-distribution}, where the small
overlap is visible for relative variations in $\bfund$ that are less than one
part in a thousand. Since carrying out a reweighting programme in the
critical region will be computationally proibitive, we reverted to an
estimate of the maximum involving a comparison of values at neighbour
simulated $\bfund$.

\begin{figure}[ht]
  \centering
  \includegraphics[width=0.7\textwidth]{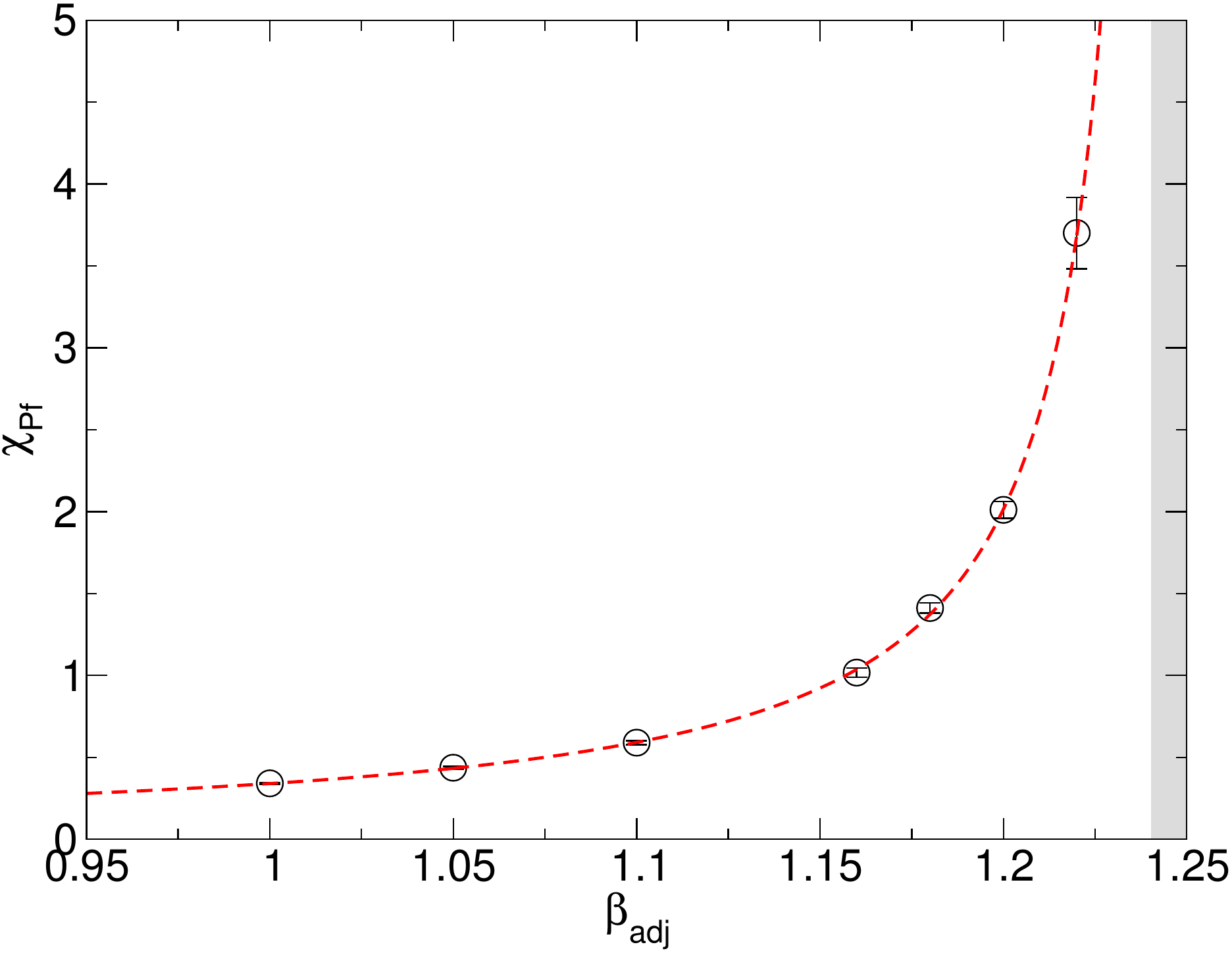}
  \caption{Scaling of the fundamental plaquette susceptibility maximum values
    for different $\badj$ couplings toward the location of the bulk
    transition end--point. The estimated $\badj^{\rm (crit)}$ from
    a fit with Eq.~(\ref{eq:fit-susc-max}) is
    highlighted by the grey shaded region on the right.}
  \label{fig:plaquette-susc-scaling}
\end{figure}

Our results show that by
approaching $\badj \approx 1.25$ from below, the maximum of the
susceptibility increases, at fixed $L^4$ volume. Its scaling form can
be fitted by
\begin{equation}
  \label{eq:fit-susc-max}
  \chif^{\rm (max)} \; = \; A  \, (\badj^{\rm (crit)} - \badj)^{-\gamma}
  \ ,
\end{equation}
with $A=0.077(5)$, $\badj^{\rm (crit)} = 1.2460(38)$ and
$\gamma=1.06(5)$ ($\chi^2/{\rm dof} = 0.67$) for the data $\badj =
1.05$-$1.22$, at $L=20$. Our numerical data for  $\chif^{\rm (max)} $
and their best fit are shown in
Fig.~\ref{fig:plaquette-susc-scaling}. $\chif^{\rm (max)}$ is the
integrated plaquette--plaquette correlation
function. The compatibility within 1.2 standard deviations of the
fitted value $\gamma = 1.06(5)$ with  $\gamma = 1$ (predicted by the
mean--field theory) gives a first hint that the model could be in the
universality class of the 4d Gaussian model, whose critical properties
are described by the mean--field approximation. The fitted $\badj^{\rm
  (crit)}$ provides another estimate of the end--point location, which is
compatible with the value obtained from $\Delta p_{\rm fund}$ . The
stability of the fit is checked by changing the number of points
included and the fitted parameters are summarised in
Tab.~\ref{tab:fit-susc-badj}. 

\begin{figure}[ht]
  \centering
  \includegraphics[width=0.7\textwidth]{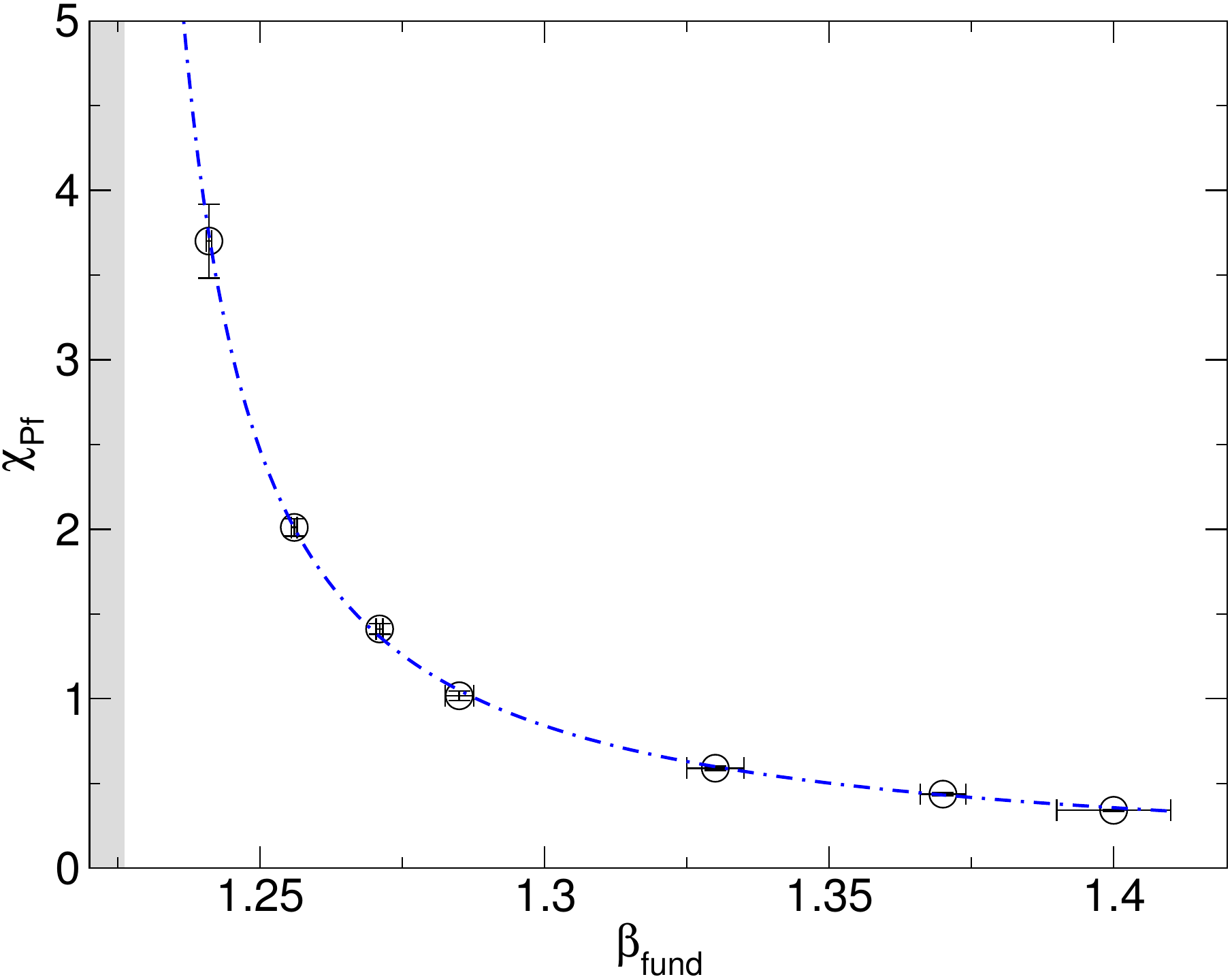}
  \caption{Scaling of the fundamental plaquette susceptibility maximum values
    for different $\bfund$ couplings toward the location of the bulk
    transition end--point. The estimated $\bfund^{\rm (crit)}$ from
    a fit with Eq.~(\ref{eq:fit-susc-max-2}) is
    highlighted by the grey shaded region on the left.}
  \label{fig:plaquette-susc-scaling-2}
\end{figure}

One can also study the scaling as a function of $\bfund$. Since our
calculations were designed for the scaling in terms of $\badj$, our
resolution for this analysis is not optimal, because $\bfund$ is
measured rather than inputed. This affects most the 
$\bfund$ values far from the critical point, and in particular $\bfund=1.40-1.42$.
Hence, we use this analysis only as a consistency check of general
scaling behaviour. In terms of
$\bfund$, the scaling form of $\chif^{\rm(max)}$ is given by 
\begin{equation}
  \label{eq:fit-susc-max-2}
  \chif^{\rm (max)} \; = \; A  \, (\bfund -\bfund^{\rm (crit)} )^{-\gamma}
  \ .
\end{equation}
Our best fit (displayed with the numerical data in
Fig.~\ref{fig:plaquette-susc-scaling-2}) gives $A=0.060(6)$,
$\bfund^{\rm (crit)} = 1.2229(31)$ and 
$\gamma=1.03(6)$ ($\chi^2/{\rm dof} = 1.41$) for the data at $\bfund =
1.241$-$1.37$ and $L=20$. Tab.~\ref{tab:fit-susc-bfund}, which reports
the values of the fitted parameters for various choices of the fit
range, suggests that the extracted value of the critical coupling and of the
critical exponent remain stable across the different fits. Finally, despite
using the same notation, we remark that the exponents $\gamma$ in
Eqs.~(\ref{eq:fit-susc-max})~and~(\ref{eq:fit-susc-max-2})
need not to be the same. The fact that the measured values are
consistent within errors is a likely indication that both the $\badj$
and the $\bfund$ directions have a non--zero projection along the
dominant direction of the renormalisation group flow at the relevant
infrared fixed point. 


\section{Spectrum measurements}
\label{sec:spectr-meas}

\begin{figure}[hb]
  \centering
  \includegraphics[width=0.7\textwidth]{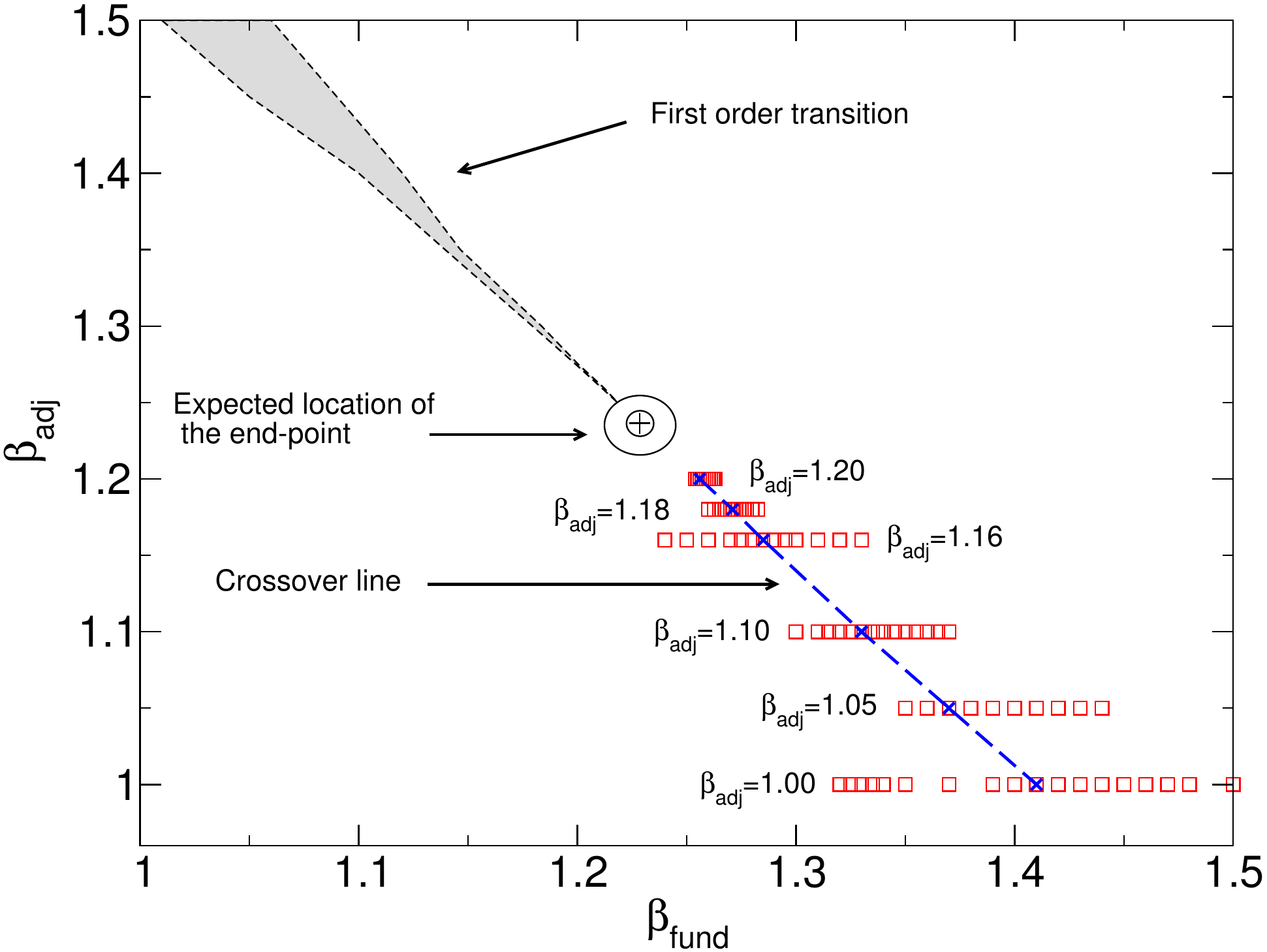}
  \caption{Location of the bulk phase transition (grey area delimited
    by thin dashed black
    lines) and of the points where we measured the spectrum of the
    theory (red squares). The estimated location of the bulk transition end--point
    is indicated by concentric circles. The thick dashed blue line joins
    the points where $\chif$ reaches its maximum and defines a crossover
    region.}
  \label{fig:simulated-points}
\end{figure}
The properties of the spectrum and the scaling of different masses as
the critical surface of a fixed point is approached give important information on the
low--energy dynamics of the theory. In the following we explain our
analysis of the low--lying spectrum in the crossover
region. Since this is the first study of this kind, we focused on
controlling possible sources of systematic errors such as
autocorrelations and finite--size effects. The aim is to extract the
light glueball spectrum in the thermodynamic limit and to study
the scaling properties of ratios of masses while approaching the
end--point in a controlled manner, e.g. along a 
specified trajectory in the bare parameter space.

We simulate the SU(2) Yang--Mills theory at six different adjoint
couplings $\badj = 1.00, \ 1.05, \ 1.10,$ $1.16, \ 1.18, \ 1.20$. For
each of them we simulate a range of fundamental couplings $\bfund$
such that both the strong coupling and the weak coupling limits are
investigated. The location of the simulated points in the
two--dimensional space of the couplings is plotted in
Fig.~\ref{fig:simulated-points}, together with the location of the bulk
phase transition and its estimated end--point
(cfr. Sect.~\ref{sec:phase-diagram}). We use several lattice volumes ranging from $6^3 \times
12$ to $48^3 \times 48$ in order to try and reach the thermodynamic
limit. Thanks to the study of the local observables needed for the
phase diagram, we were able to estimate the autocorrelation times and
we noted a drastic increase of them for $1.16 \le \badj \le
1.20$ (cfr. Tab.~\ref{tab:susept-all}). Therefore, different measurements are separated by $N_{\tau}$
gauge updates to reduce autocorrelations, with $N_{\tau}=250$ at
$\badj=1.16$ up to $N_{\tau}=800$ at $\badj=1.20$. Large statistics
ensembles with $N_{\rm meas} \ge 10000$ measurements allow us to
extract masses of gluon bound states with great accuracy. A
variational ansatz using a large basis of interpolating
operators is employed to extract the ground state and excited state
masses in different symmetry channels as described in
Ref.~\cite{Lucini:2010nv}.

\begin{figure}[ht]
  \centering
  \includegraphics[width=0.7\textwidth]{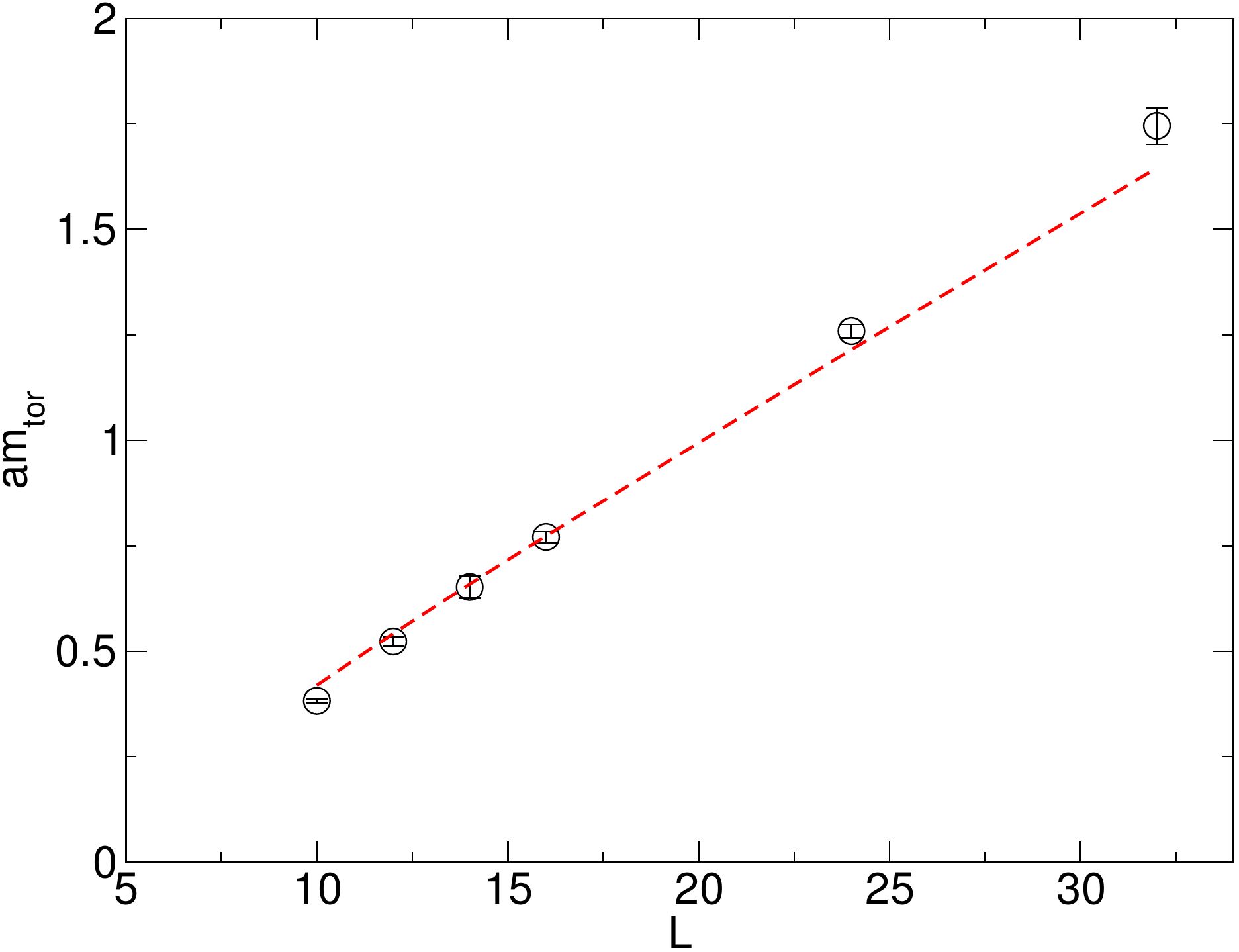}
  \caption{Dependence of the torelon mass $am_{\rm tor}$ on its length $L$ at
    fixed $\badj=1.20$, $\bfund=1.259$. The dashed line is
    Eq.~\eqref{eq:sigma} with $\Mst=0.2290$, obtained from $am_{\rm
      tor}(L)$ at $L=16$. Torelon masses larger than the cutoff are not considered
  as reliable estimates.}
  \label{fig:torelon-ba120-bf1259}
\end{figure}
\begin{figure}[ht]
  \centering
  \includegraphics[width=0.7\textwidth]{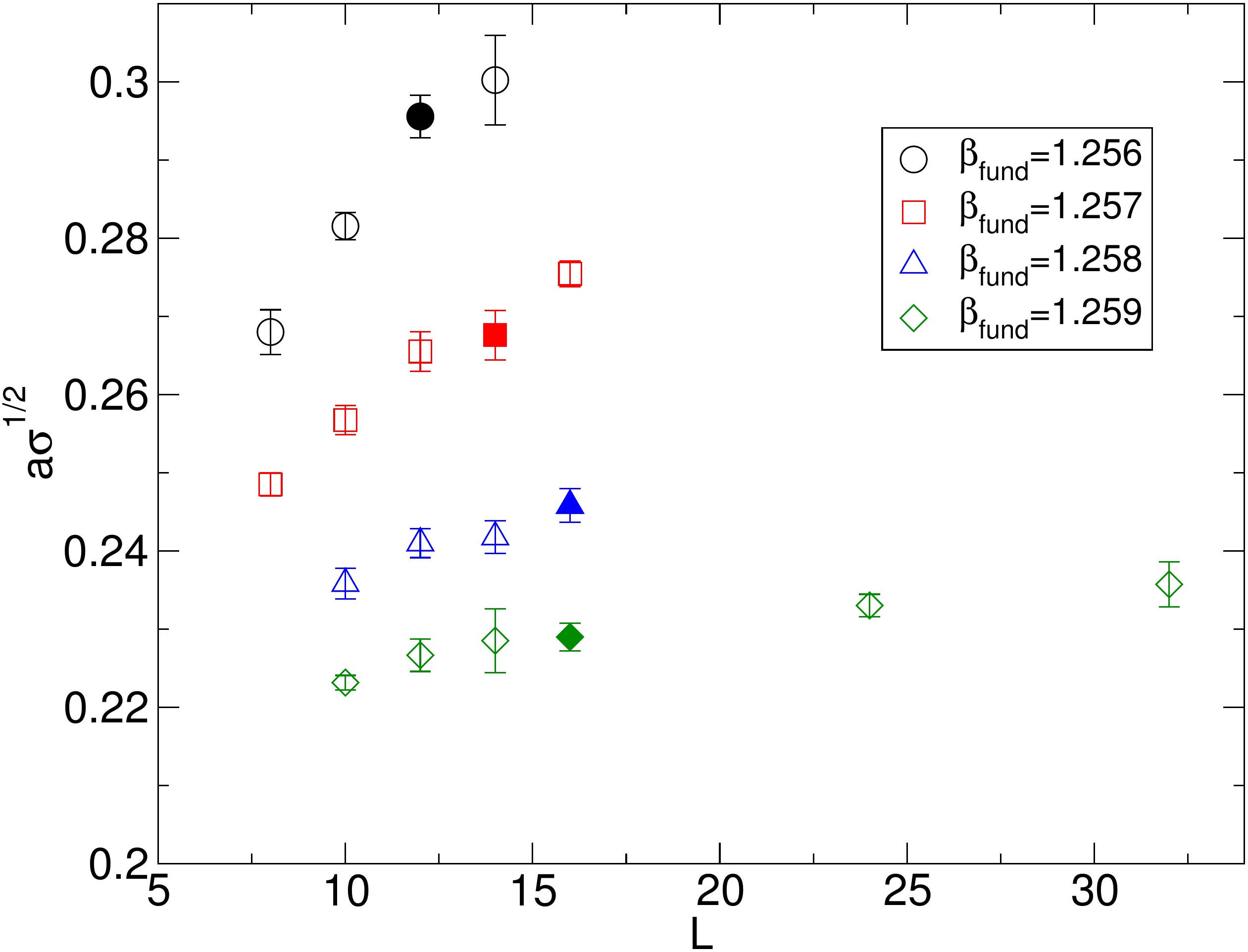}
  \caption{Dependence of the string tension $\Mst$ on the length of the
    spatial torelons $L$ at fixed $\badj=1.20$ and for different
    fundamental couplings. The filled points identify the results used
  to estimate the infinite volume limit of $\Mst$.}
  \label{fig:infinite-volume-string-ba120}
\end{figure}
\begin{figure}[ht]
  \centering
  \includegraphics[width=0.7\textwidth]{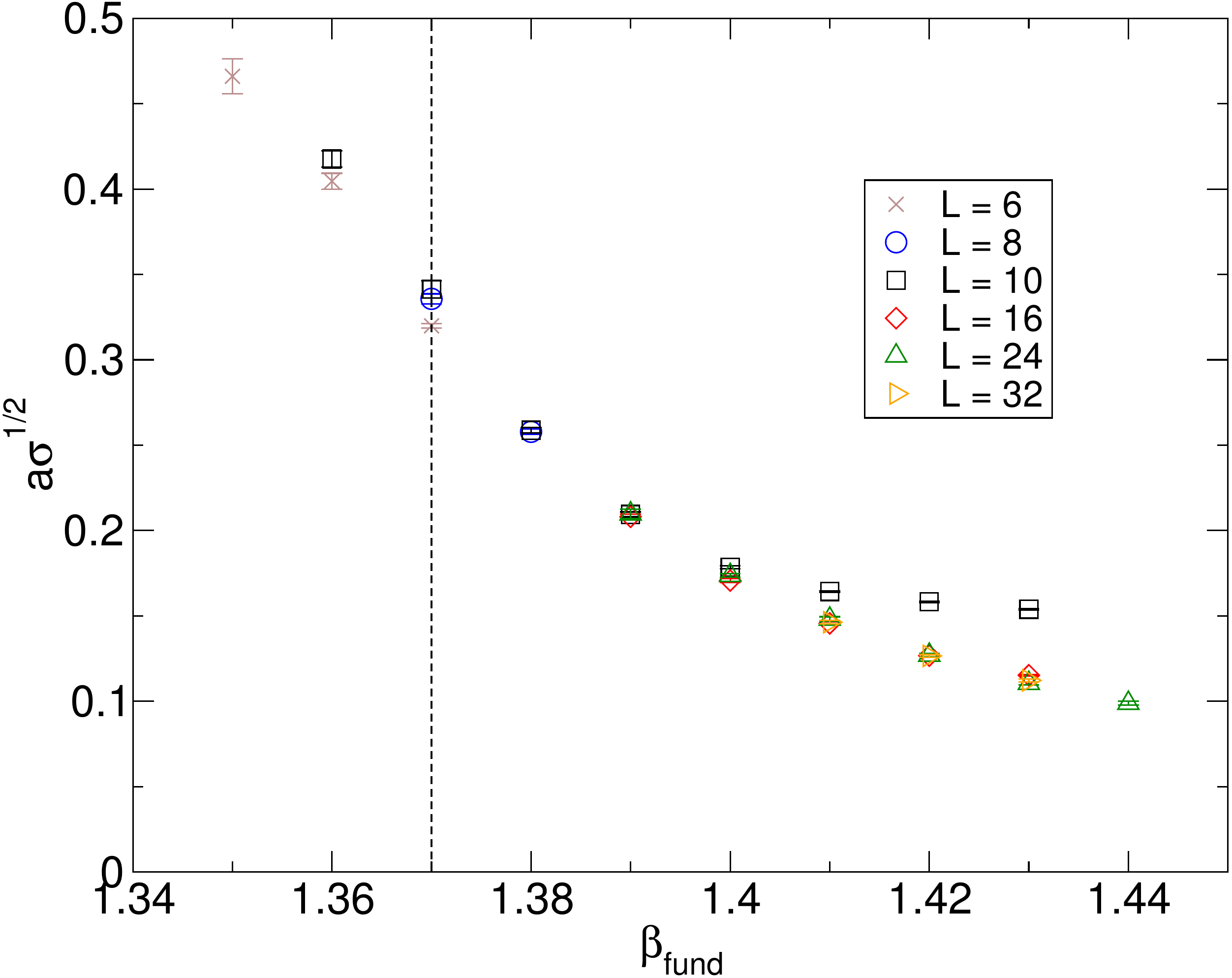}
  \caption{String tension measured from several different volumes
    $L=6$-$32$ at $\badj=1.05$. At weaker coupling $\bfund > 1.39$
    larger volumes are needed to keep finite--volume effects under
    control. The dashed vertical line indicates the approximate position of the
    maximum in $\chif$.}
  \label{fig:string-tension-ba105}
\end{figure}
First of all, we extracted the string tension $\Mst$, which was then
used to set the overall
scale. This measure of the dynamical mass gap is extracted from long
spatial Polyakov loop correlators. The asymptotic large--time behaviour of 
these correlators is governed by the lightest torelon state. Assuming
that a confining flux tube with massless modes binds a static
quark-antiquark pair, the mass of the lightest torelon $am_{\rm tor}$
can be used to obtain the string tension according to the ansatz
\begin{equation}
  \label{eq:sigma}
  am_{\rm tor}(L)  =  a^2\sigma L - \frac{\pi}{3L} -
  \frac{\pi^2}{18L^3}\frac{1}{a^2\sigma}
  \ .
\end{equation}
The validity of the above equation is checked {\it a posteriori} by
comparing the extracted string tension at various sizes $L$ and by evaluating
the size of the subleading finite $L$ correction $c_2 =
\pi^2/(18L^3a^2\sigma)$ with respect to the leading one $c_1 = -\pi/(3L)$. This
procedure is illustrated on a typical set of data in
Fig.~\ref{fig:torelon-ba120-bf1259}, where the string tension is not
fitted, but rather extracted using Eq.~(\ref{eq:sigma}) 
and the data point $am_{\rm tor}(L)$ at $L = 16$. We observe that, when the
loop is too short (i.e. such that  $am_{\rm tor}(L) < 0.5$), the asymptotic
formula relating string tension, length of the loop and torelon mass fails to
describe the latter. On  the other hand, extracting masses that are
above the lattice cutoff  is subject to systematic errors. There is an
intermediate region of masses, $0.5 \le am_{\rm tor}(L) \le 1.0$, that
are correctly described by Eq.~(\ref{eq:sigma}). For the numerical
value at the highest simulated $L$ in the regime of validity of Eq.~(\ref{eq:sigma}) (which in
this case is the result at $L=16$) one gets $c_2/c_1 \simeq 0.039$. Besides
justifying our procedure, these results give support to
the existence of confining fluxes below the critical point (see also
Fig.~\ref{fig:infinite-volume-string-ba120}, showing an example of
the variation of $\Mst$ as a function of $L$ and $\bfund$ at fixed
$\badj$). Our results show that significant finite--size
effects are absent when $La\sqrt{\sigma} > 3$, which we satisfied
in our simulations using large spatial volumes for the smallest values of
$\Mst$. Indeed, in the explored range of couplings, the string
tension can change by a factor of $5$ and, whereas small $L\sim 8-10$
volumes are sufficient at stronger couplings, larger ones are needed
towards weak coupling. The situation is shown in
Fig.~\ref{fig:string-tension-ba105} and it is representative of all
the simulated $\badj$ points.

We can estimate the infinite--volume limit of this observable in the
following way: when the two largest simulated volumes at each point
give consistent results within two standard deviations, we take the largest volume
result as an estimate of the thermodynamic limit (provided $am_{\rm
  tor}$ is below the cutoff). When the aforementioned criterion
is not fulfilled, we do not give an infinite--volume
estimate. However, if a single volume simulation is available, we
still report it in plots that show results on various volumes.
We use the same approach for estimating the large volume limit also for the
other spectral observables studied in this section.

\begin{figure}[ht]
  \centering
  \includegraphics[width=0.7\textwidth]{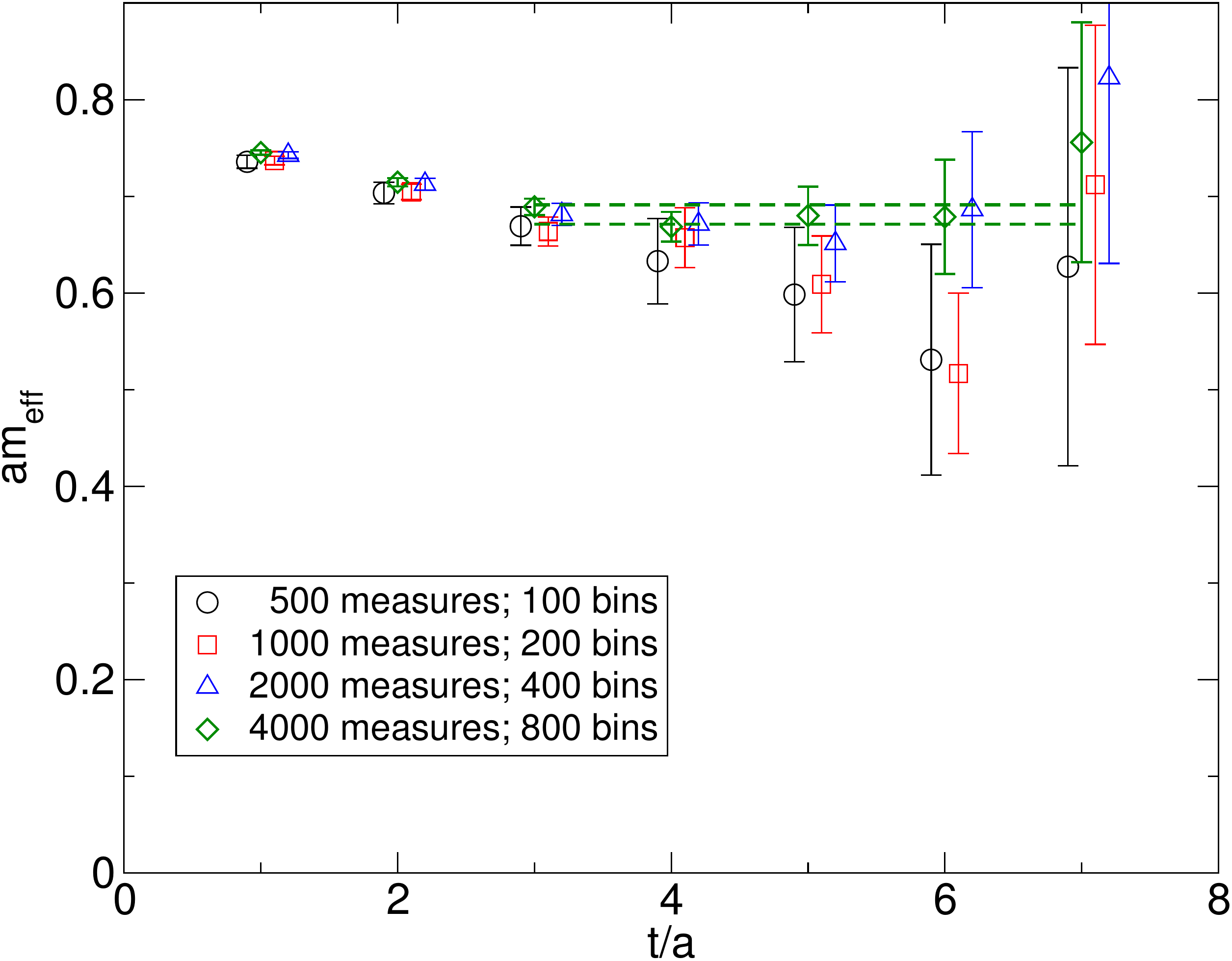}
  \caption{Effective $am_{\rm tor}$ at $\badj=1.20$ and
    $\bfund=1.256$ for $L=10$, $T=20$. The statistics
    is doubled until a plateaux is clearly
    identified. The dashed green lines indicate the $1$-$\sigma$ contour
    of the fitted effective mass between $t=3$ and $t=7$. Data points
    are horizontally shifted for clarity.}
  \label{fig:plateaux-stat}
\end{figure}

Another
important point that we mentioned in Sect.~\ref{sec:phase-diagram} is
the increase in the number of measurements that are
needed closer to $\badj \approx 1.25$ due to large autocorrelation
times. For example, in
Fig.~\ref{fig:plateaux-stat} we show that almost a tenfold increase in
statistics is needed to reduce the systematic error in the
identification of the effective mass plateaux for $am_{\rm tor}$ at

\begin{figure}[ht]
  \centering
  \includegraphics[width=0.7\textwidth]{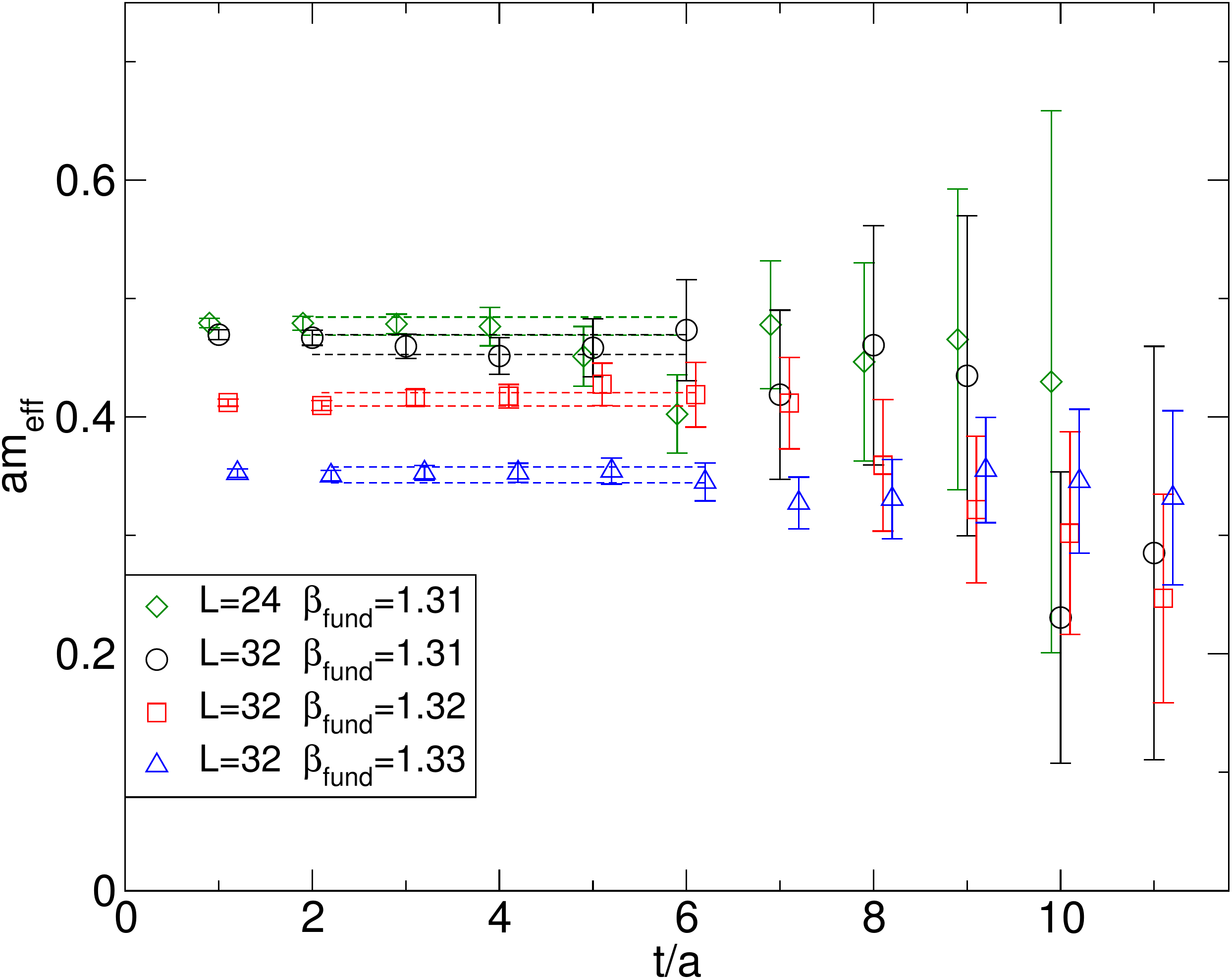}
  \caption{Scalar glueball effective mass at $\badj=1.16$ and for
    three values of $\bfund$. A comparison between $L=24$ and $L=32$
    results for $\bfund=1.31$ is also shown. The $1$-$\sigma$ contour
    of the fitted masses is plotted on top of the respective fitted
    points. Data points are horizontally shifted for clarity.}
  \label{fig:plateaux-scalar}
\end{figure}
\begin{figure}[ht]
  \centering
  \includegraphics[width=0.7\textwidth]{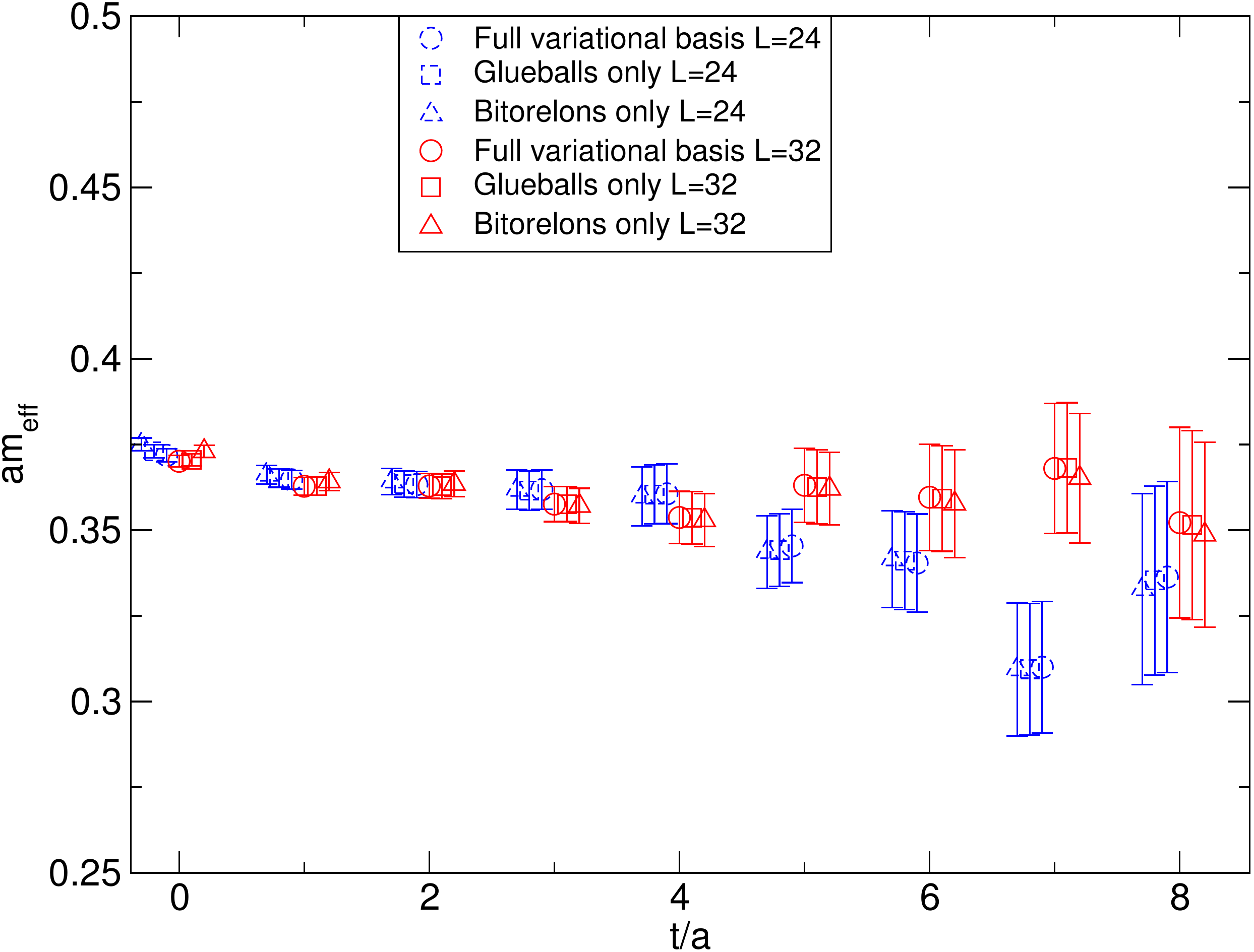}
  \caption{Effective mass of the scalar ground state obtained using
    different variational basis. Two volumes are compared at
    $\bfund=1.259$, $\badj=1.20$. No significant difference is
    present. Points are shifted for clarity.}
  \label{fig:variational-ba120-bf1259}
\end{figure}
\begin{figure}[ht]
  \centering
  \includegraphics[width=0.7\textwidth]{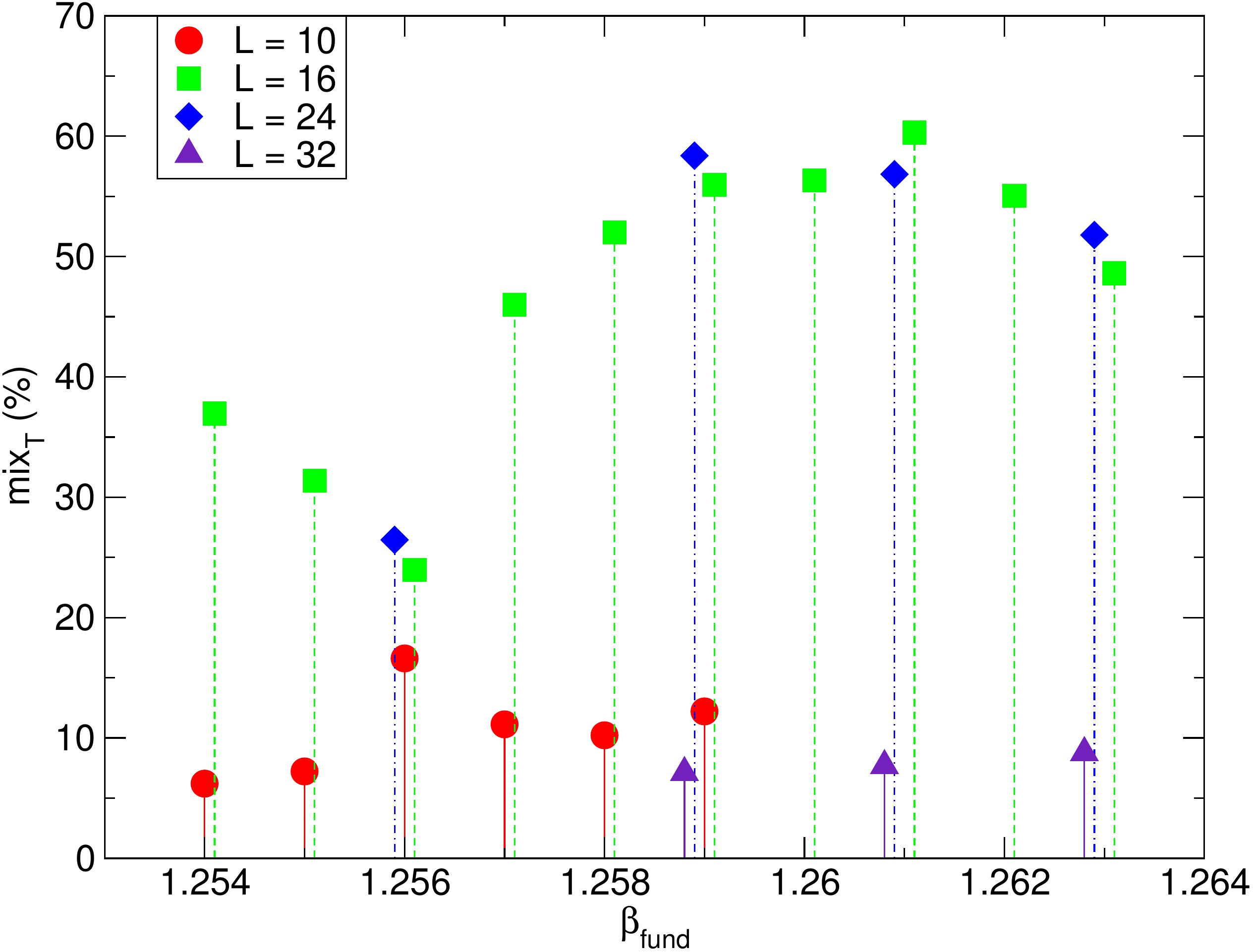}
  \caption{Relative contribution of the bi--torelon operators to the
    scalar ground state for several fundamental couplings at
    $\badj=1.20$. On the largest volume, this contribution drops
    significantly. Values at different volumes are shifted horizontally
  for clarity.}
  \label{fig:projection-ba120}
\end{figure}
Our second spectral observable is the mass of the scalar glueball
state $\Msc$. In order to measure this mass, the vacuum subtracted
 correlators of smeared Wilson loops, symmetrised to have $0^{++}$
 quantum numbers, have been inserted in a variational basis for
a generalised eigenvalue problem. In addition, in order to identify
 finite--size artefacts, a different type of scalar operators made by
 symmetrised Polyakov loops  winding in opposite directions around the
 spatial torus has been used in the same variational calculation. This
 second operator set couples mainly to bi--torelon excitations, which
 are suppressed in the large volume limit and can be used to identify
 these spurious contaminations of the spectrum in the scalar channel.
For further technical details, we refer the reader to Ref.~\cite{Lucini:2010nv}. 

The scalar glueball mass is reliably estimated thanks to the large
variational operator basis used in our calculation, which allows us to
obtain long and robust
effective mass plateaux. Fig.~\ref{fig:plateaux-scalar} provides an example
of effective masses for a large $32^4$ lattice at fixed $\badj$ for three $\bfund$
values, with a comparison with results from a smaller $24^3 \times
32$ for one value of $\bfund$. For larger $\bfund$ values, 
we used bigger volumes in order to check explicitly for finite--size
effects. Moreover, the contributions of spurious states has been
investigated by looking at the extracted spectrum using the different
kind of operators described above. Surprisingly, for some values of the
couplings we have noticed a large $\mathcal{O}(50\%)$ contribution of
the bi--torelon operators to the ground state; a variational analysis
containing only Wilson loop operators or, separately, only Polyakov
loop operators, turned out to give the same results for the ground
state masses. An example of the effective mass
plot obtained from such different variational operator bases is shown in
Fig.~\ref{fig:variational-ba120-bf1259}. This confirms that
bi--torelon operators can give 
sizeable contributions to correlators used to extract the scalar
ground state mass. However, by using larger lattices we could confirm
that the contribution of these operators dropped down to
less than $10\%$, as expected. This is clearly depicted in
Fig.~\ref{fig:projection-ba120}, where the relative bi--torelon
operators contribution to the ground state is shown at $\badj=1.20$
and for several volumes. In our computation, care has been taken in
choosing the lattice size in such a way that bi--torelon contamination
in the scalar spectrum is negligible. Results for the infinite volume
limit of $\Mst$ and $\Msc$ at various fundamental and adjoint
couplings are included in Tab.~\ref{tab:ba100} to~\ref{tab:ba120}.  

The strategy applied for the extraction of
 the tensor glueball mass $\Mtn$ is similar to that used for
$\Msc$. Here, the lattice operators are symmetrised to
project only onto the $E$ irreducible representation of the cubic
group which is subduced from the spin $2$ of the full continuum
rotational symmetry. We remark that for some of the
$\badj$ couplings we could not reliably estimate the thermodynamic
limit of $\Mtn$ due to somewhat larger finite--size effects. It is
known that this channel is more difficult to extract due to the
heavier mass of its ground state. Results in the thermodynamic limit
are summarised in Tab.~\ref{tab:ba100-ba105-tensor} for this observable.

\begin{figure}[ht]
  \centering
  \includegraphics[width=0.7\textwidth]{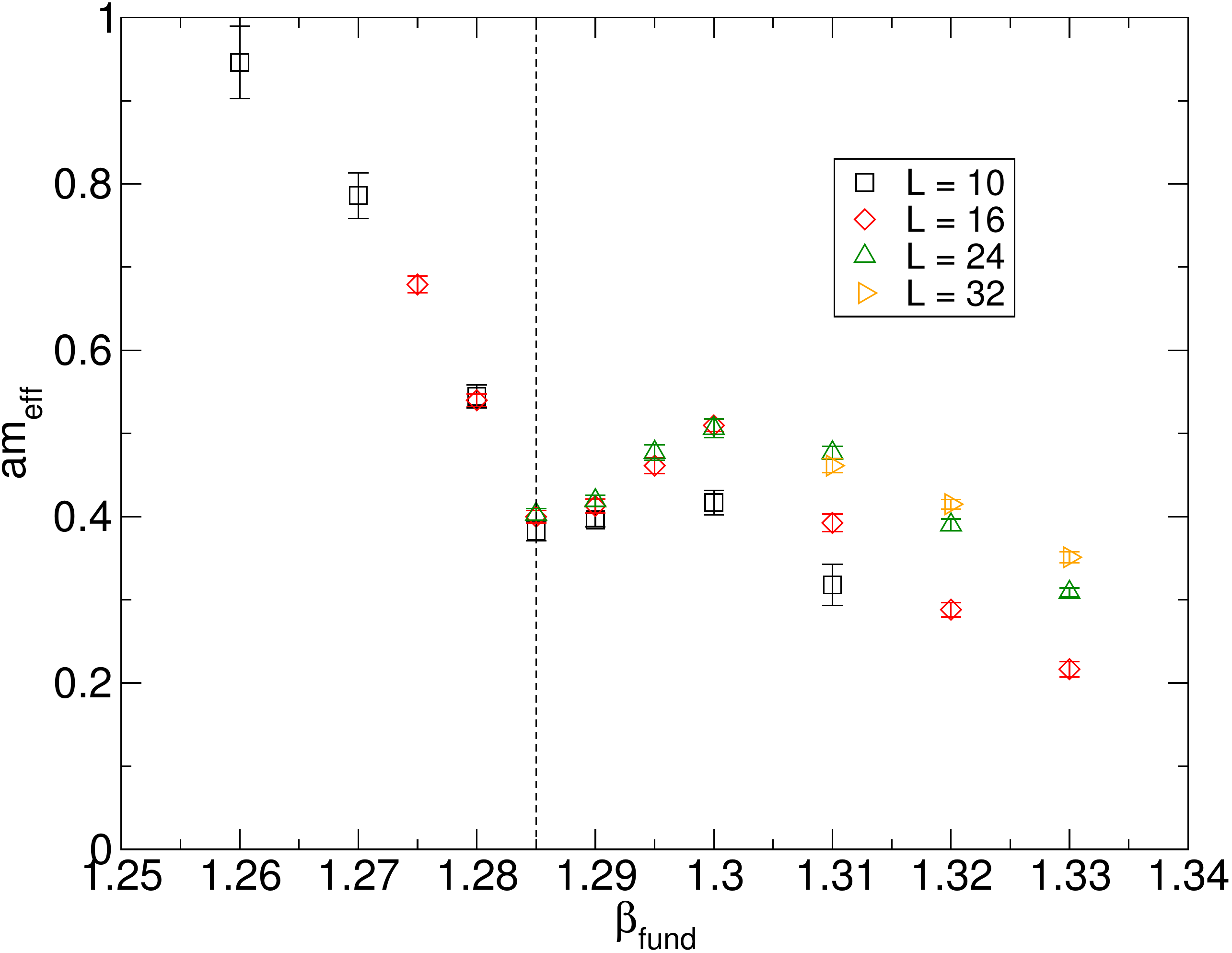}
  \caption{Scalar glueball mass at $\badj=1.16$ for a wide range of
    fundamental couplings. Contrary to the monotonicity of $\Mst$ in
    the same range of $\bfund$, $\Msc$ develops a dip in the crossover region,
    before raising and decreasing again in the weak coupling
    limit. The dashed vertical line indicates the approximate position of the
    maximum in $\chif$.}
  \label{fig:scalar-ba116}
\end{figure}
\begin{figure}[ht]
  \centering
  \includegraphics[width=0.7\textwidth]{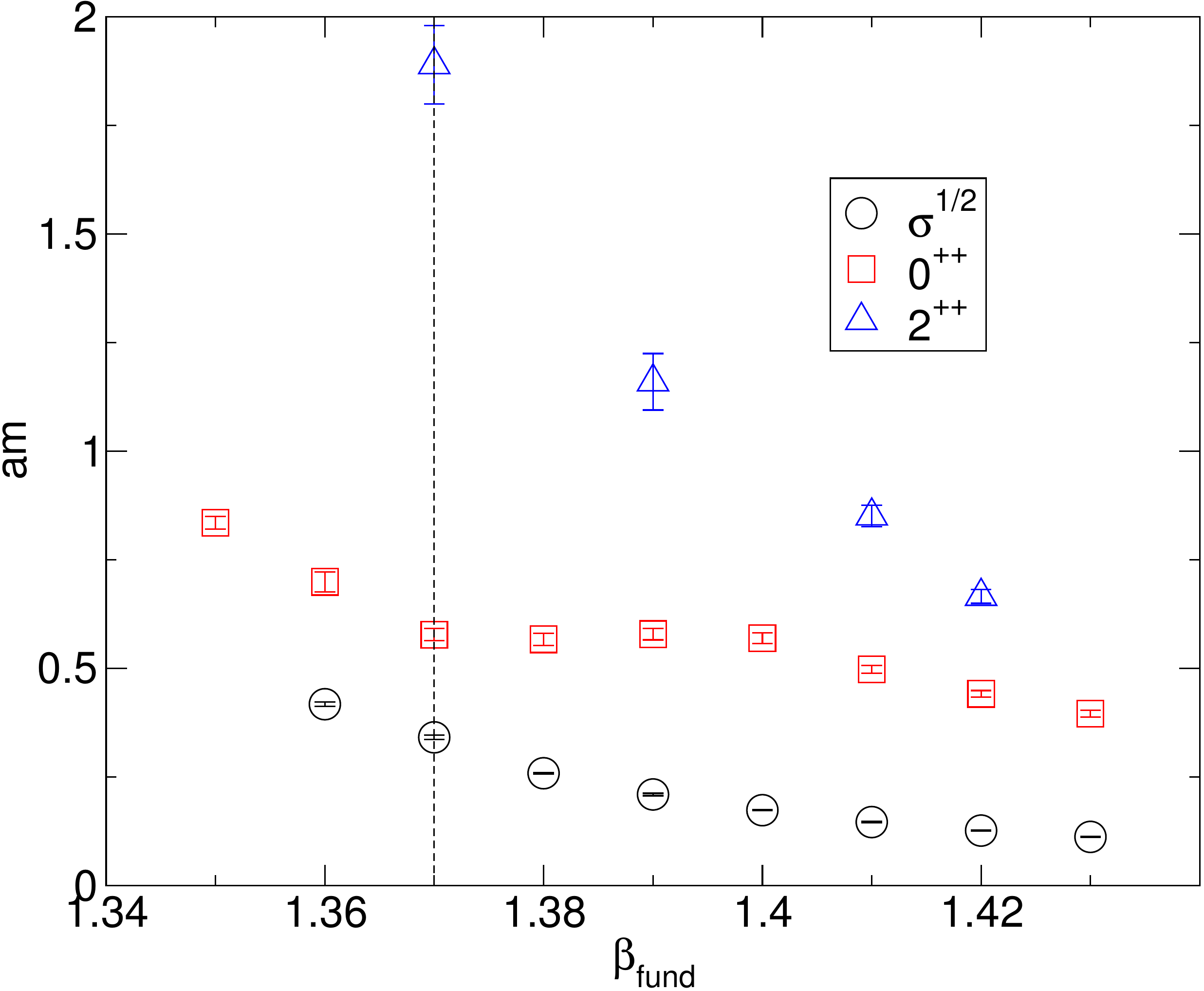}
  \caption{The measured low--lying spectrum of $\Mst$, $\Msc$ and
    $\Mtn$ at $\badj=1.05$ in the infinite--volume limit. The dashed
    vertical line indicates the approximate position of the 
    maximum in $\chif$.}
  \label{fig:spectrum-ba105}
\end{figure}

\section{Scaling properties}
\label{sec:scal-prop}
 Let us move to the description of the features of the extracted
spectrum, when the relavant sources of finite--size effects are taken
into account. The most interesting feature
of the scalar channel spectrum is the presence of a slight dip for a
certain region of $\bfund$ couplings around the crossover region,
where $\chif$ reaches its maximum value. This dip becomes more
pronounced and at its bottom the mass value gets lighter as we increase $\badj$
towards the transition end--point. At $\badj=1.00$ the dip is still
only a mild plateaux that $\Msc$ reaches before starting decreasing
again towards the weak coupling region. However, at $\badj=1.16$
$\Msc$ drops dramatically, to form the dip shown in
Fig.~\ref{fig:scalar-ba116}. A similar situation has been found at
$\badj=1.18$ and $1.20$. It is important to recall that such a
behaviour is absent in both $\Mst$ and $\Mtn$, which smoothly decrease
as functions of $\bfund$. A situation where the
infinite--volume limit has been estimated is shown in
Fig.~\ref{fig:spectrum-ba105}.

\begin{figure}[ht]
  \centering
  \includegraphics[width=0.7\textwidth]{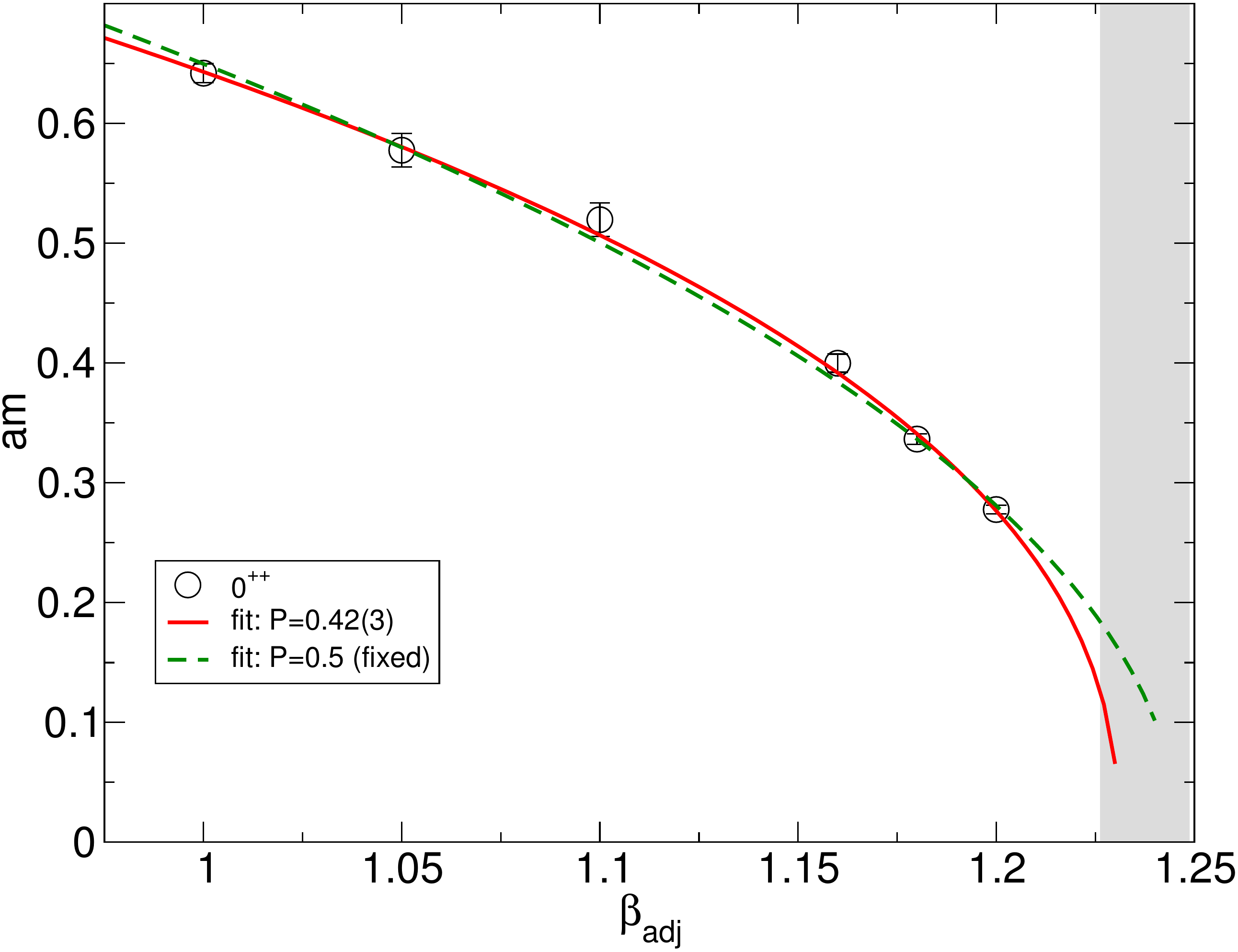}
  \caption{$\Msc$ for different values of $\badj$ and on the trajectory
    defined by the maxima of $\chif$. The fitting function from
    Eq.~(\ref{eq:fit-am-scal}) is used to represent the data. The shaded
  grey area comprises the values of the critical point coming from the
two different fits in the plot.} 
  \label{fig:scaling}
\end{figure}

\begin{figure}[ht]
  \centering
  \includegraphics[width=0.7\textwidth]{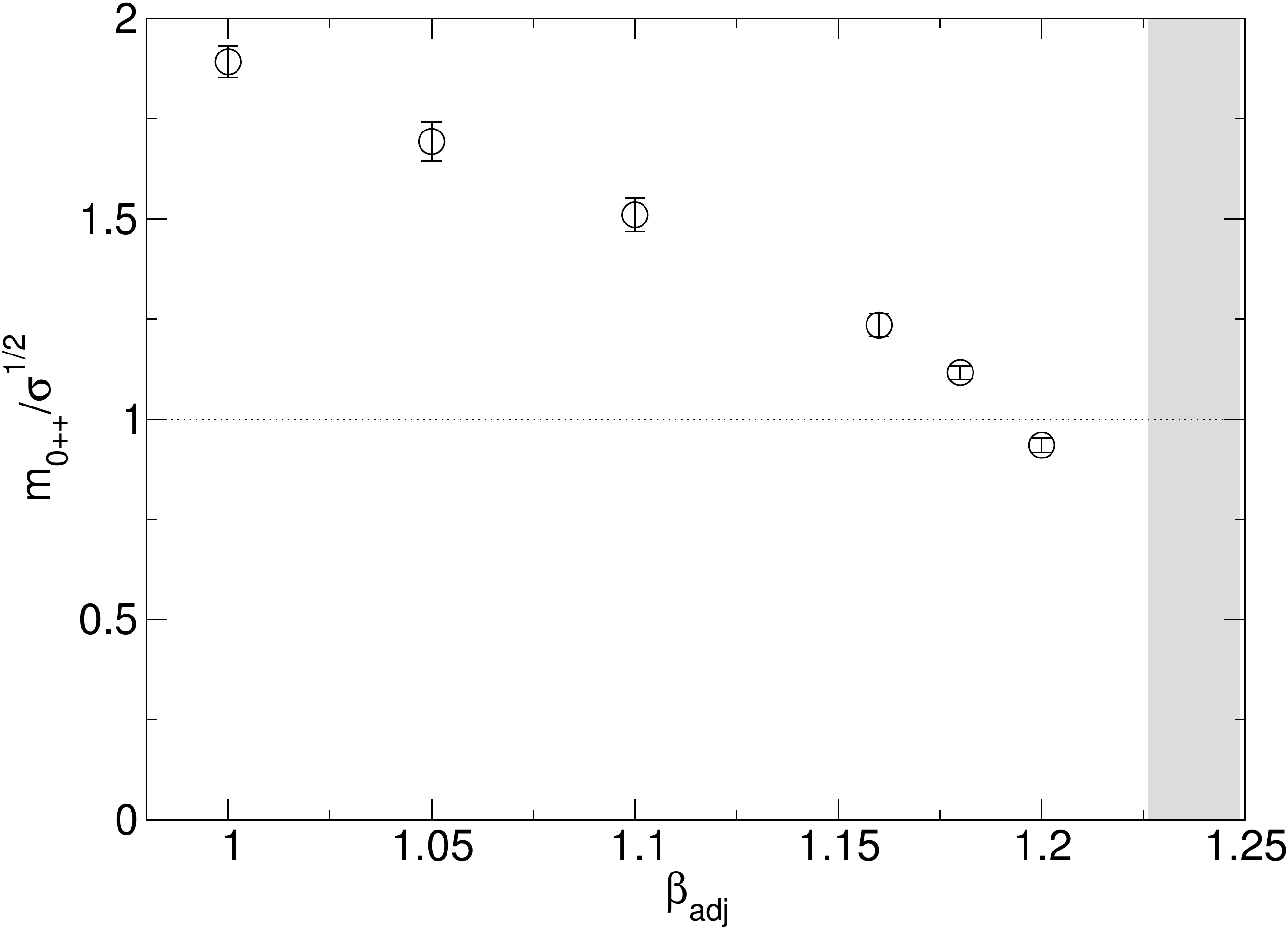}
  \caption{The ratio between $\Msc$ and $\Mst$ for different values of
    $\badj$ and on the trajectory defined by the maxima of $\chif$. The ratio
    decreases below one when the bulk transition end--point is
    approached ($\Delta p_{\rm fund}=0$). The shaded grey area
    indicates the estimated location of the critical point obtained by
    the scaling analysis (Eq.~(\ref{eq:fit-am-scal})).}
  \label{fig:ratio}
\end{figure}

In a neighbourhood of a second order phase transition point the
light lattice degrees of freedom that become massless at the critical
point define an effective long--distance continuum theory. As the
critical point is approached, their mass goes to zero according to
some scaling exponents that characterise the dynamics at large distances. For our system, only
$\Msc$ seems to become light at the end--point of the first order
line. In order to investigate its approach to the end--point, we fit the
measured $\Msc$ using the parameterisation
\begin{equation}
  \label{eq:fit-am-scal}
  \Msc \; = \; A  \, (\badj^{\rm (crit)} - \badj)^P
  \ ,
\end{equation}
which is inspired by the scaling of the correlation length $\xi$ near a
critical point: $\xi = \xi_0 \left| T/T_c - 1 \right|^{-\nu}$.
The critical exponent
$\nu$ is $0.5$ for the 4d Gaussian model (mean--field theory). By fitting
the three free parameters using all the 
available data, we obtain $A=1.19(5)$, $\badj^{\rm (crit)} = 1.2308(59)$
and $P=0.42(3)$ with a good $\chi^2/{\rm dof} = 1.07$. We also fitted
the data keeping a fixed $P=0.5$ to test the mean--field hypothesis and
we obtain $A=1.31(5)$, $\badj^{\rm (crit)} = 1.2455(26)$ with a worsened
$\chi^2/{\rm dof} = 1.97$. Both fits are compared to the data in
Fig.~\ref{fig:scaling}. Taken at face value, our results suggest that the mean--field
scaling is not ruled out. Indeed if we exclude the point at $\badj =
1.00$, the value of $P$ gets closer to the mean--field prediction
(Tab.~\ref{tab:fitm0pp-badj}).  As shown in Tab.~\ref{tab:fitm0pp-fund}, a
similar analysis of the scaling of $\Msc$ in terms of $\bfund$ gives compatible results.
To resolve the issue of whether the system is described by the
mean--field theory, one would need to go closer to the end--point,
which is currently computationally proibitive for the resources at our
disposal. The $m_{0^{++}}/\sqrt{\sigma}$ ratio is shown in
Fig.~\ref{fig:ratio}.   

\section{Conclusions}
\label{sect:conclusions}
Motivated by the need to better understand possible roles of lattice
artefacts in investigations of gauge theories in the (near-)conformal
regime, we have studied a SU(2) pure gauge theory with a
modified lattice action with couplings to both the
fundamental and the adjoint plaquettes. This theory, which is related to
SU(2) gauge theory with two Dirac flavours in the adjoint
representation in the limit of large bare fermion mass,
is known to have a bulk phase transition with an end--point relatively
close to the fundamental coupling axis. The controversial nature of
this end--point is resolved and our estimates for its location are
summarised in Tab.~\ref{tab:fitcritpoint}. Using our improved gluon
spectroscopy techniques~\cite{Lucini:2010nv}, we measured the string
tension, the scalar glueball mass and the tensor glueball mass. We
studied their scaling properties when the end--point is approached
along a controlled trajectory that follows the peaks of the
fundamental plaquette susceptibility. To our knowledge, this is the
first systematic study of the
gluonic spectrum in this model. For this reason, we carefully checked
that we are reasonably free from finite--size effects and (mostly thanks to the
simulation algorithm used in this work) the autocorrelation time of
our observables was kept under control. The spectrum extrapolated to infinite
volume shows a non--constant $m_{0^{++}}/\sqrt{\sigma}$ ratio when
approaching the end--point in a controlled manner (see Fig.~\ref{fig:ratio}). In particular,
since the $0^{++}$ state is the only light degree of freedom near the
end--point and the scaling is marginally compatible with being
described by the critical exponents of the 4d Gaussian model, it is
conceivable that the corresponding effective theory be a 
scalar theory described by the mean-field approximation. This
behaviour is in contrast with the infrared dynamics of SU(2) gauge theory with two adjoint
Dirac fermions, where such a ratio is driven to a constant by a
conformal fixed point and is consistent with the continuum SU(2) Yang--Mills value
$m_{0^{++}}/\sqrt{\sigma} \sim 3.7$. Therefore, we can reasonably
conclude that the observed spectral signals
of near-conformality in SU(2) gauge theory with two adjoint Dirac
fermions are not affected by the second order phase transition point
of the related gauge system with mixed fundamental-adjoint action. It
would be instructive to perform a similar analysis for gauge theories
with fermions in the symmetric or antisymmetric representation and
$N_c \ge 4$, for which the stability of fluxes of higher ${\cal
  N}$-lity in pure gauge (see e.g.~\cite{Lucini:2000qp}) could create a more
complicated phase structure in the effective theory at large mass. 

\acknowledgments{
We would like to thank Philippe de Forcrand, Claudio Bonati and Guido
Cossu for fruitful comments and discussions. We are indebted with Urs
Heller for guidance on the algorithm used in our
simulations. B.L. acknowledges financial support from the Royal
Society (grant UF09003) and STFC (grant ST/G000506/1) and the hospitality of the
Aspen Center for Physics during the final stage of this work, which
allowed him to discuss various aspects of the project with the
participants to the workshop {\it Lattice Gauge Theory in the LHC
  Era}, and in particular with R. Brower, A. Hasenfratz, Y. Meurice
and T. Tomboulis. E.R. was funded by a SUPA Prize Studentship and a
FY2012 JSPS Postdoctoral Fellowship for Foreign Researchers
(short-term). The simulations discussed in this work have been
performed on a cluster partially funded by STFC and by the Royal
Society, on systems made available to us by HPC Wales and on the HPCC
Plymouth cluster facilities at Plymouth University.  
}

\bibliographystyle{JHEP}
\bibliography{su2fa}

\begin{table}[H]
  \centering
  \begin{tabular}{c|c|c|c}
    $L$ & $\badj$ & $\bfund$ & $\Delta p_{\rm fund}$ \\
    \hline
    8 & 1.50  &1.04    & 0.25(1) \\
    8 & 1.45  &1.075  & 0.22(1) \\
    8 & 1.40  &1.110  & 0.19(1) \\
    8 & 1.35  &1.144  & 0.161(1) \\
    12 &1.30  & 1.1815& 0.117(3) \\
    16 & 1.275& 1.1999& 0.085(1) \\
    20 &1.26  & 1.2109& 0.0592(1) \\
    40 &1.25  & 1.2183$^*$& 0.02(1)$^*$ \\
  \end{tabular}
  \caption{The estimated location of the hysteresis centre at different
    $\badj$ values. Each value $\Delta p_{\rm fund}$ is measured on
    a volume $L^4$, which is the minimum one needed to discern the two
    metastable states of the first order transition. The errors on
    $\Delta p_{\rm fund}$ are estimated by comparing its value on all
    the simulated points in the hysteresis region. However, the starred
    point comes from a single simulation, and the error comes from the
    difficulty of estimating the expectation values $\VEV{{\rm
        Plaq}_{F,1}}$ and $\VEV{{\rm Plaq}_{F,2}}$ in the presence of
    long autocorrelation times.}
\label{tab:delta-p-f}
\end{table}

\begin{table}[H]
  \centering
  \begin{tabular}{c|c|c|c|c|c}
    $\badj$ & $\bfund$ & $\chif^{\rm (max)}$ & $\tau_{\rm f}$ & $\chia^{\rm (max)}$ & $\tau_{\rm a}$\\
    \hline
    1.00  &1.40(1) & 0.3415(55) & 45(2) & 0.2102(30) & 40(2)\\
$^*$1.00 & 1.41(1) & 0.3307(58) & 52(3) & 0.2261(37)& 47(5) \\
    1.05 &1.370(4) & 0.4366(82) & 72(4) & 0.2834(49) & 66(4)\\
    1.10 &1.330(5) & 0.590(12) & 86(6) & 0.3658(77) & 81(5) \\
    1.16 &1.2850(25) & 1.017(29)  & 175(17) &  0.591(17)  & 168(16)\\
    1.18 &1.2710(6) & 1.412(32) & 276(34) & 0.835(18) & 244(16) \\
    1.20 &1.2560(5) & 2.011(51) & 401(59) & 1.175(29) & 366(30) \\
    1.22 & 1.2410(5) & 3.70(22) &  623(113) & 2.14(13) &  617(112) \\ 
  \end{tabular}
  \caption{The maximum of the fundamental and adjoint plaquette
    susceptibilities $\chif^{\rm (max)}$, $\chia^{\rm (max)}$
    and the integrated autocorrelation time of the fundamental and
    adjoint plaquettes
    $\tau_{\rm f}$, $\tau_{\rm a}$ for different adjoint couplings $\badj$ and at fixed
    volume $L=20$. The statistical errors come from a jackknife procedure 
    with bins of $3 \tau$ measures, while the error on $\bfund$
    is estimated by the distance between neighbouring simulated
    points. The two lines at $\badj = 1.00$ are due to the fact that
    both $\bfund = 1.40$ and $\bfund = 1.41$ give compatible values
    for the measured maximum of the susceptibility, which indicates
    that the real maximum falls in between the two simulated
    $\bfund$. The  starred point at $\badj = 1.00$ was used for the calculation of the
    spectrum at $\badj=1.00$, while the other point at the same value
    of $\badj$ was used for the susceptibility analyses. The fact that
    at fixed $\badj$ different (close) values of $\bfund$ are suitable
    for a scaling analysis of different observables is due to the
    mildness of the crossover at the boundary of the critical region.}
\label{tab:susept-all}
\end{table}

\begin{table}[H]
  \centering
  \begin{tabular}{c|c|c|c|c}
    range  $\badj$ & $\badj^{\rm (crit)}$ & A & $\gamma$ & $\chi^2/{\rm dof}$\\
    \hline
    1.00-1.20 & 1.2453(55) & 0.078(5) & 1.05(6) & 0.66 \\
    1.00-1.22 & 1.2459(28) & 0.077(3) & 1.06(3) & 0.50 \\
    1.05-1.20 & 1.2451(87) & 0.078(10) & 1.05(10) & 1.00 \\
    {\bf 1.05-1.22} & {\bf 1.2460(38)} & {\bf 0.077(5)} & {\bf 1.06(5)} &
    {\bf 0.67} \\
    1.00-1.22$^*$ & 1.2485(39) & 0.072(4) & 1.10(5) & 0.79 \\
  \end{tabular}
  \caption{Fit results for $\chif^{\rm (max)}$ according to the formula $\chif^{\rm (max)}\; = \; A  \,
    (\badj^{\rm (crit)} - \badj)^{-\gamma}$. The starred value uses the
    starred point in Tab.~\ref{tab:susept-all}. Boldfaced values are
    used in the text.}
\label{tab:fit-susc-badj}
\end{table}

\begin{table}[H]
  \centering
  \begin{tabular}{c|c|c|c|c}
    range  $\bfund$ & $\bfund^{\rm (crit)}$ & A & $\gamma$ & $\chi^2/{\rm dof}$\\
    \hline
    1.241-1.40 & 1.2206(34) & 0.053(5) & 1.09(6) & 1.69 \\
    {\bf 1.241-1.37} &  {\bf 1.2229(31)} &  {\bf 0.060(6)} &  {\bf 1.03(6)} &  {\bf 1.41} \\
    1.256-1.40 & 1.2201(72) & 0.053(8) & 1.10(11) & 2.24 \\
    1.256-1.37 & 1.2256(68) & 0.063(12) & 0.99(11) & 1.96 \\
    1.256-1.41$^*$ & 1.2220(26) & 0.057(41) & 1.05(4) & 1.16 \\
  \end{tabular}
  \caption{Fit results for $\chif^{\rm (max)}$ according to the
    formula $\chif^{\rm (max)}\; = \; A  \, (\bfund -
    \bfund^{\rm (crit)} )^{-\gamma}$. The starred value uses the
    starred point in Tab.~\ref{tab:susept-all}. Boldfaced values are
    used in the text.}
\label{tab:fit-susc-bfund}
\end{table}

\begin{table}[H]
  \centering
    \begin{tabular}{c||c|c|c||c|c|c}
      \multicolumn{7}{c}{$\badj=1.00$}\\
      \hline \hline
      $\bfund$ & $L$ & $\Msc$ & $t_i$-$t_f$ & $L$ & $\Mst$ & $t_i$-$t_f$ \\
      \hline
      1.39 & 10 & 0.838(34)  & 2-5 &  10 & 0.4554(96) & 1-4 \\
      1.40 & 16 & 0.726(20)  & 2-5 &  10 & 0.3958(36) & 1-4 \\ 
      1.41 & 24 & 0.635(14)  & 2-6 &  16 & 0.3393(57) & 1-4 \\
      1.42 & 24 & 0.615(15)  & 2-6 &  24 & 0.2770(42) & 1-4 \\
      1.43 & 24 & 0.625(11)  & 2-5 &  24 & 0.2323(13) & 1-4 \\
      1.44 & 24 & 0.591(12)  & 2-5 &  24 & 0.1942(8)   & 1-5 \\
      1.45 & 32 & 0.5781(77) & 1-6 &  32 & 0.1635(15) & 2-6 \\
      1.46 & 32 & 0.4953(91) & 2-6 &  32 & 0.1456(19) & 3-8 \\
      1.47 & 32 & 0.4593(88) & 2-6 &  32 & 0.1263(14) & 3-10 \\
   \end{tabular}
\caption{Values for $\Msc$ and $\Mst$ on the lattices used as an
  estimate of the infinite volume limit for different $\bfund$ at
  $\badj=1.00$. The time interval $t_i$-$t_f$ indicates the fitting
  window used for the reported values.}
\label{tab:ba100}
\end{table}

\begin{table}[H]
  \centering
    \begin{tabular}{c||c|c|c||c|c|c}
      \multicolumn{7}{c}{$\badj=1.05$}\\
      \hline \hline
      $\bfund$ & $L$ & $\Msc$ & $t_i$-$t_f$ & $L$ & $\Mst$ & $t_i$-$t_f$ \\
      \hline
      1.35 & 10 & 0.835(14) & 1-5  & 6 & 0.466(10) $^*$ & 1-4 \\
      1.36 & 10 & 0.699(23) & 1-7  & 10 & 0.4177(47) &  1-4 \\
      1.37 & 16 & 0.578(14) & 2-6  & 10 & 0.3413(50)  & 2-5 \\
      1.38 & 16 & 0.567(14) & 2-6  & 10 & 0.2588(15)  & 2-6 \\
      1.39 & 24 & 0.579(13) & 2-7  & 24 & 0.2098(28)   & 2-6 \\ 
      1.40 & 24 & 0.569(12) & 2-6  & 24 & 0.1737(11)   & 2-5 \\ 
      1.41 & 32 & 0.4977(92) & 2-6 & 32 & 0.1463(10) & 2-6 \\
      1.42 & 32 & 0.4414(77) & 2-6 & 32 & 0.1265(7) & 2-6 \\
      1.43 & 32 & 0.3961(80) & 2-6 & 32 & 0.1123(9) & 3-9 \\
   \end{tabular}
\caption{Same as Tab.~\ref{tab:ba100} for $\badj=1.05$. The stars $^*$ indicate quantities for which only a single volume simulation is available.}
\label{tab:ba105}
\end{table}

\begin{table}[H]
  \centering
    \begin{tabular}{c||c|c|c||c|c|c}
      \multicolumn{7}{c}{$\badj=1.10$}\\
      \hline \hline
      $\bfund$ & $L$ & $\Msc$ & $t_i$-$t_f$ & $L$ & $\Mst$ & $t_i$-$t_f$ \\
      \hline
      1.32  & 10& 0.716(2)$^*$  & 2-5 & 10 & 0.4331(67) & 1-4 \\
      1.325 & 10 & 0.591(1)$^*$ & 1-6 &  10 & 0.3961(44)$^*$ & 1-4 \\
      1.330 & 16 & 0.519(14) & 2-6 &  10 & 0.3442(24) & 1-4\\
      1.335 & 16 & 0.488(12) & 2-6 &  16 & 0.2954(27) & 1-4\\
      1.340 & 16 & 0.504(12) & 2-6 &  16 & 0.2507(14) & 1-4\\
      1.345 & 24 & 0.547(12) & 2-7 &  24 & 0.2178(11) & 1-4\\
      1.350 & 32 & 0.553(11) & 2-7 &  24 & 0.1895(18) & 2-7\\
      1.360 & 32 & 0.542(11) & 2-7 &  32 & 0.1566(25) & 3-6\\
      1.370 & 32 & 0.4555(92) & 2-6 &  32 & 0.1275(7) & 2-6\\
   \end{tabular}
\caption{Same as Tab.~\ref{tab:ba100} for $\badj=1.10$. The stars $^*$ indicate quantities for which only a single volume simulation is available.}
\label{tab:ba110}
\end{table}

\begin{table}[H]
  \centering
    \begin{tabular}{c||c|c|c||c|c|c}
      \multicolumn{7}{c}{$\badj=1.16$}\\
      \hline \hline
      $\bfund$ & $L$ & $\Msc$ & $t_i$-$t_f$ & $L$ & $\Mst$ & $t_i$-$t_f$ \\
      \hline
      1.275 & 16 & 0.679(10)$^*$  & 1-5 & -- & -- & -- \\
      1.28  & 16 & 0.5399(76)  & 1-6 & 10 & 0.417(21) & 2-5 \\
      1.285 & 24 & 0.4028(69)  & 2-7 & 12 & 0.3254(52) & 2-5 \\
      1.29  & 24 & 0.4192(66)  & 2-7 & 16 & 0.2485(30) & 2-5 \\
      1.295 & 24 & 0.4772(93)  & 2-7 & 24 & 0.2012(23) & 2-7 \\
      1.30  & 24 & 0.506(11)  & 2-7 &  24 & 0.1736(12) & 2-6 \\
      1.31  & 32 & 0.4612(85)  & 2-6 & 32 & 0.1348(8) & 2-8 \\
      1.32  & 32 & 0.4149(57)  & 2-6 & 32 & 0.1117(9) & 3-9 \\
      1.33 &  --     &  --      &  --    & 32 & 0.0958(5) & 3-9 \\
   \end{tabular}
\caption{Same as Tab.~\ref{tab:ba100} for $\badj=1.16$. The stars $^*$ indicate quantities for which only a single volume simulation is available.}
\label{tab:ba116}
\end{table}

\begin{table}[H]
   \centering
   \begin{tabular}{c||c|c|c||c|c|c}
      \multicolumn{7}{c}{$\badj=1.18$}\\
      \hline \hline
      $\bfund$ & $L$ & $\Msc$ & $t_i$-$t_f$ & $L$ & $\Mst$ &   $t_i$-$t_f$ \\
      \hline
      1.2600 & 16 & 0.693(11) & 1-6 & -- & -- & -- \\
      1.2625 & 16 & 0.645(16) & 2-5 & -- & -- & -- \\
      1.2650 & 16 & 0.5213(73) & 1-6 & -- & -- & -- \\
      1.2675 & 16 & 0.4458(93) & 2-6 & 10 & 0.403(15)$^*$ & 2-4 \\
      1.2700 & 32 & 0.3472(51) & 2-7 & 10 & 0.3224(45)$^*$ & 2-4 \\
      1.2710 & 32 & 0.3343(49) & 2-7 & 12 & 0.3015(25) & 2-5 \\
      1.2725 & 24 & 0.3439(46) & 2-7 & 12 & 0.2698(30) & 2-5 \\
      1.2750 & 24 & 0.3975(60) & 2-7 & 24 & 0.2384(15) & 1-6 \\
      1.2775 & 24 & 0.4440(71) & 2-7 & 24 & 0.2102(9) & 1-6 \\
      1.2800 & 24 & 0.4874(83) & 2-7 & 24 & 0.1869(13) & 2-5 \\
      1.2825 & 24 & 0.5135(95)$^*$ & 2-7 & 24 & 0.1724(10)$^*$ & 2-5 \\
   \end{tabular}
\caption{Same as Tab.~\ref{tab:ba100} for $\badj=1.18$. The stars $^*$ indicate quantities for which only a single volume simulation is available.}
\label{tab:ba118}
\end{table}

\begin{table}[H]
   \centering
   \begin{tabular}{c||c|c|c||c|c|c}
      \multicolumn{7}{c}{$\badj=1.20$}\\
      \hline \hline
      $\bfund$ & $L$ & $\Msc$ & $t_i$-$t_f$ & $L$ & $\Mst$ & $t_i$-$t_f$ \\
      \hline
      1.254 & 16 & 0.3424(55) & 2-6 & 10  & 0.3499(68)  & 2-4 \\
      1.255 & 16 & 0.2963(43) & 2-7 & 10  & 0.3245(49)  & 2-4 \\
      1.256 & 48 & 0.2771(42) &3-10 & 12  & 0.2971(39)  & 2-5 \\
      1.257 & 16 & 0.3008(49) & 2-7 & 14  & 0.2676(32)  & 2-6 \\
      1.258 & 16 & 0.3189(44) & 2-6 & 16  & 0.2458(22)  & 2-4 \\
      1.259 & 32 & 0.3595(54) & 2-8 & 16  & 0.2290(18)  & 2-5 \\
      1.261 & 32 & 0.4122(73) & 2-7 & 32  & 0.2095(14)  & 1-3 \\
      1.263 & 32 & 0.4522(76) & 2-7 & 32  & 0.1896(9)  & 1-4 \\
   \end{tabular}
\caption{Same as Tab.~\ref{tab:ba100} for $\badj=1.20$.}
\label{tab:ba120}
\end{table}

\begin{table}[H]
   \centering
   \begin{tabular}{c|c|c|c||c|c|c|c}
      \multicolumn{4}{c||}{$\badj=1.00$} & \multicolumn{4}{c}{$\badj=1.05$}\\
      \hline \hline
      $\bfund$ & $L$ & $\Mtn$ & $t_i$-$t_f$ & $\bfund$ & $L$ & $\Mtn$ & $t_i$-$t_f$ \\
      \hline
      1.39 & 10 & 2.56(37) & 1-3 & 1.37 & 32 & 1.888(94) & 1-4 \\
      1.40 & 10 & 2.39(19) & 1-3 & 1.38 & 16 & 1.455(49)$^*$  & 1-4 \\
      1.41 & 16 & 1.92(11) & 1-5 & 1.39 & 32 & 1.160(65) & 2-5 \\
      1.42 & 24 & 1.581(50) & 1-5 & 1.40 & 24 & 0.910(29)$^*$  & 2-5 \\
      1.43 & 24 & 1.237(26) & 1-4 & 1.41 & 32 & 0.851(25) & 2-5 \\
      1.44 & 24 & 1.048(17) & 1-4 & 1.42 & 32 & 0.666(16) & 2-6 \\
      1.45 & 32 & 0.909(31) & 2-6 & -- & -- & -- & -- \\
      1.46 & 32 & 0.732(20) & 2-6 & -- & -- & -- & -- \\
      1.47 & 32 & 0.687(14) & 2-6 & -- & -- & -- & -- \\
   \end{tabular}
\caption{Values for $\Mtn$ on the lattices used as an
  estimate of the infinite volume limit for different $\bfund$ at
  $\badj=1.00$ and $\badj=1.05$. The time interval $t_i$-$t_f$ indicates the fitting
  window used for the reported values. The stars $^*$ indicate quantities for which only a single volume simulation is available.}
\label{tab:ba100-ba105-tensor}
\end{table}

\begin{table}[H]
  \centering
  \begin{tabular}{c|c|c|c|c}
    range  $\badj$ & $\badj^{\rm (crit)}$ & A & $P$ & $\chi^2/{\rm dof}$\\
    \hline
    1.00-1.20 & 1.2308(59) & 1.19(5) & 0.42(3) & 1.07 \\
    1.05-1.20 & 1.2330(96) & 1.23(13) & 0.44(7) & 1.48 \\
    1.00-1.20 & 1.2455(26) & 1.31(2) & 0.5 & 1.97 \\
    1.05-1.20 & 1.2420(28) & 1.36(3) & 0.5 & 1.37 \\
  \end{tabular}
  \caption{Fit results for $\Msc$ according to the formula $\Msc\; = \; A  \, (\badj^{\rm (crit)} -
    \badj)^P$. In the last two lines, $P$ is kept fixed to 0.5.}
\label{tab:fitm0pp-badj}
\end{table}

\begin{table}[H]
  \centering
  \begin{tabular}{c|c|c|c|c}
    range  $\bfund$ & $\bfund^{\rm (crit)}$ & A & $P$ & $\chi^2/{\rm dof}$\\
    \hline
    1.256-1.41 & 1.2346(47) & 1.30(8) & 0.40(4) & 1.64 \\
    1.256-1.37 & 1.2326(77) & 1.37(20) & 0.43(7) & 2.23 \\
    1.256-1.41 & 1.2210(25) & 1.50(3) & 0.5 & 3.14 \\
    1.256-1.37 & 1.2246(26) & 1.57(5) & 0.5 & 2.10 \\
  \end{tabular}
  \caption{Fit results for for $\Msc$ according to the formula $\Msc \; = \; A  \, (\bfund - \bfund^{\rm
      (crit)} )^P$. In the last two lines, $P$ is kept fixed to 0.5.}
\label{tab:fitm0pp-fund}
\end{table}

\begin{table}[H]
  \centering
  \begin{tabular}{c|c|c}
    method & $\badj^{\rm (crit)}$ & $\bfund^{\rm (crit)}$ \\
    \hline
    latent heat (Eq.~\eqref{eq:latent-heat}) & 1.22 - 1.25 & --\\
    $\chif^{\rm (max)}$ scaling (Eq.~\eqref{eq:fit-susc-max}) & 1.2460(38) & 1.2229(31)\\
    $\Msc$ scaling (Eq.~\eqref{eq:fit-am-scal}) & 1.2308(59) & 1.2346(47)\\
                                  with fixed exponent ($P=0.5$) & 1.2455(26) & 1.2210(25) \\
  \end{tabular}
  \caption{Summary of the different estimates of the critical $(\bfund,\badj)$ values described in
    the text.}
\label{tab:fitcritpoint}
\end{table}

\end{document}